\documentclass[twocolumn,showpacs,nofootinbib,superscriptaddress,prd,aps,10pt,floatfix]{revtex4-2}

\usepackage[nointlimits]{amsmath}
\usepackage{hyperref}
\usepackage{amsfonts}
\usepackage{amssymb}
\usepackage{amsmath}
\usepackage{amsthm}
\usepackage{latexsym}
\usepackage{graphicx}

\usepackage{color}
\usepackage{layout}
\usepackage{orcidlink}
\usepackage[english]{babel}
\hypersetup{
 colorlinks   = true, %Colours links instead of boxes
 urlcolor     = blue, %Colour for external hyperlinks
 linkcolor    = blue, %Colour of internal links
 citecolor   = blue %Colour of citations
 }

\usepackage{wrapfig}

\def\LL{\mathcal{L}}
\def\x{\mathbf{x}}

\def\vv{\mathbf{v}}
\def\a{\mathbf{a}}

\def\eps{\varepsilon}
\def\OO{{\cal O}}
\def\VV{{\cal V}}
\def\eps{\varepsilon}

\allowdisplaybreaks

\begin{document}

\title{Cosmological and lunar laser ranging constraints on evolving dark energy\\
in a nonminimally coupled curvature-matter gravity model}

\author{Riccardo March \orcidlink{0000-0003-3144-7537}}
\email{r.march@iac.cnr.it}
\affiliation{Istituto per le Applicazioni del Calcolo, CNR, Via dei Taurini 19, 00185 Roma, Italy}
\affiliation{INFN - Laboratori Nazionali di Frascati (LNF), Via E. Fermi 40, Frascati 00044 Roma, Italy}

\author{Miguel Barroso Varela \orcidlink{0009-0006-9844-7661}}
\email{up201907272@edu.fc.up.pt}
\affiliation{Departamento de F\'isica e Astronomia,\\Faculdade de Ci\^encias da Universidade do Porto, Rua do Campo Alegre 687, 4169-007 , Porto, Portugal}
\affiliation{Centro de F\'isica das Universidades do Minho e do Porto, Rua do Campo Alegre 687, 4169-007, Porto, Portugal}

\author{Orfeu Bertolami \orcidlink{0000-0002-7672-0560}}
\email{orfeu.bertolami@fc.up.pt}
\affiliation{Departamento de F\'isica e Astronomia,\\Faculdade de Ci\^encias da Universidade do Porto, Rua do Campo Alegre 687, 4169-007 , Porto, Portugal}
\affiliation{Centro de F\'isica das Universidades do Minho e do Porto, Rua do Campo Alegre 687, 4169-007, Porto, Portugal}

\author{\\Giada Bargiacchi \orcidlink{0000-0002-0167-8935}}
\email{Giada.Bargiacchi@lnf.infn.it}
\affiliation{INFN - Laboratori Nazionali di Frascati (LNF), Via E. Fermi 40, Frascati 00044 Roma, Italy}

\author{Marco Muccino \orcidlink{0000-0002-2234-9225}}
\email{marco.muccino@unicam.it}
\affiliation{Universit\'a di Camerino, via Madonna delle Carceri, Camerino, 62032, Italy}
\affiliation{Al-Farabi Kazakh National University, Al-Farabi av. 71, 050040 Almaty, Kazakhstan}
\affiliation{ICRANet, Piazza della Repubblica 10, Pescara, 65122, Italy}

\author{Simone Dell'Agnello \orcidlink{0000-0002-0691-8213}}
\email{simone.dellagnello@lnf.infn.it}
\affiliation{INFN - Laboratori Nazionali di Frascati (LNF), Via E. Fermi 40, Frascati 00044 Roma, Italy}

\date{\today}

\begin{abstract}
We analyze a cosmological solution to the field equations of a modified gravity model where curvature and matter are nonminimally coupled.
The current Universe's accelerated expansion is driven by a cosmological constant while the impact of the nonminimal coupling on the expansion history is recast as an effective equation of state for evolving dark energy. The model is analyzed under a tracking solution that follows the minimum of the effective potential for a scalar field that captures the modified theory's effects. We determine the conditions for the existence of this minimum and for the validity of the tracking solution.
Cosmological constraints on the parameters of the model are obtained by resorting to recent outcomes of data from the DESI collaboration in combination with the Pantheon+ and Dark Energy Survey supernovae compilations, which give compatible results that point to the presence of a dynamical behavior for dark energy. The gravity model violates the equivalence principle since it gives rise to a fifth force that implies the Earth and Moon fall differently towards the Sun. 
The cosmological constraints are intersected with limits resulting from a test of the equivalence principle in the Earth-Moon system based on lunar laser ranging data. We find that a variety of model parameters are consistent with both of these constraints, all while producing a dynamical evolution of dark energy with similarities to that found in recent DESI results.
\end{abstract}

\maketitle

%%%%%%%%%%%%%%%%%%%%%%%%%%%%%%%%%%%%%%%%%%%%%%%%%%%%%%%%%%
\section{Introduction}
%%%%%%%%%%%%%%%%%%%%%%%%%%%%%%%%%%%%%%%%%%%%%%%%%%%%%%%%%%
The past decades have seen the emergence of an increasing number of issues on the accuracy of the standard model of cosmology ($\Lambda$CDM). The inconsistency of direct late-time and indirect early-time measurements of the  Hubble constant, commonly referred to as the Hubble tension, seems to point an inconsistency between our knowledge of the low and high-curvature Universe \cite{Agrawal:2019lmo,DiValentino:2021izs}. Furthermore, recent discoveries by the Dark Energy Spectroscopic Instrument (DESI) collaboration have revealed signs that dark energy may be evolving with time, departing from the cosmological constant which takes center stage in the $\Lambda$CDM model \cite{DESI:2024mwx,DESI:2025zgx}. The origin of such behavior may range from theories of quintessence \cite{Zlatev:1998tr,Kamenshchik:2001cp,Wolf:2024eph}, the generalized Chaplygin gas \cite{Bilic:2001cg,Gorini:2002kf,Bento:2002ps} or even interacting dark energy \cite{Li:2024qso,Giare:2024smz}. 

Another theoretical possibility is that our current theory of gravity, General Relativity (GR), may not correctly describe all physical scales. In the spirit of effective theory representations, a logical extension is to modify the Einstein-Hilbert action from its original linear dependency on the Ricci curvature scalar $R$ into a general function of this same quantity \cite{Sotiriou:2008rp,DeFelice:2010aj}. The resulting $f(R)$ theories branch into a considerable variety of models, with some impact on the inflationary mechanism at the early Universe \cite{Starobinsky:2007hu}, while others attempt to provide an explanation for the present accelerated expansion of the Universe \cite{HS}. 

Following similar methodology, one is led to consider the possibility of introducing a non-minimal coupling (NMC) between matter and curvature. A direct method for breaking the minimal coupling of GR is to insert an additional coupling $f^2(R)$ multiplying the matter field Lagrangian in the action. The NMC $f(R)$ theory was originally introduced in Ref. \cite{BBHL} in the context of modifying the geodesic equation, thus leading to an extra force that could reproduce the effect of dark matter on the rotation curves of galaxies without introducing new matter content into the theory \cite{drkmattgal,Bertolami:2011ye}. This proposal has since then been thoroughly researched in the context of the late Universe as a mechanism to mimic the effects of a cosmological constant and thus drives the accelerated expansion of the Universe \cite{curraccel}. Indeed, besides accounting for this acceleration, it can also provide a successful explanation for the Hubble tension by modifying the evolution of Cosmic Microwave Background (CMB) data from the Planck collaboration \cite{Planck:2018vyg} into the present in a way that increases the present expansion rate and thus patches together indirect and direct measurements of the expansion rate, $H_0$ \cite{BarrosoVarela:2024htf}. Other research on the theory includes the analysis of its effects on the propagation and polarizations of gravitational waves \cite{Bertolami:2017svl,Varela:2024egg}, studies on growth of large-scale structure in the Universe \cite{Nesseris:2008mq,Bertolami:2013kca,BarrosoVarela:2025mro} and the determination of its implications on inflationary dynamics along with the corresponding constraints on the theory's parameters \cite{Gomes:2016cwj,Bertolami:2022hjk,BarrosoVarela:2025yyr}.

In the present work we attempt to constrain a specific NMC gravity model by using both cosmological and Lunar Laser Ranging (LLR) data,
where the function $f^2(R)$ has a power law form $\mu R^m$.
The cosmological field equations consist in a modified Friedmann equation coupled with an equation for a suitable scalar function that captures
the effects of the nonminimal coupling. We analyze a tracking solution that follows the minimum of an effective potential for the scalar function during
the Universe's expansion. It turns out that such a tracking solution reproduces the behavior of an evolving dark energy.

The impact of NMC gravity on the dynamics of the Sun-Earth-Moon system has been investigated in Ref. \cite{MBMDeA}. The structure of the gravitational field equations for the three-body system shows that the solution of such equations exhibits a screening mechanism which is a NMC version of the so-called chameleon mechanism, which is typical of chameleon theories of gravity such as $f(R)$ gravity \cite{MBMGDeA,MBMDeA}. 
Due to screening, deviations from GR in the gravitational field outside of the three astronomical bodies are sourced by 
thin shells of mass close to the surfaces of the bodies: in the lunar crust; mainly in Earth's seawater; in the solar photosphere and in the top of the solar convection zone.
Such deviations give rise to a fifth force which is typical of chameleon theories of gravity. 
This fifth force depends on the mass density profiles in the thin shells of the three bodies, so that it depends on composition and size of the bodies. 
Consequently, the Earth and Moon fall towards the Sun with different accelerations giving rise to a violation of the weak equivalence principle (WEP). 
Constraints on the thickness of the shells, that translate into constraints on the parameters of the NMC gravity model, have been obtained in Ref. \cite{MBMDeA} by means of a test of WEP based on LLR measurements.

In previous papers \cite{curraccel,BarrosoVarela:2024htf}, where NMC gravity has been applied to cosmology as a mechanism to mimic the effects of a cosmological constant and thus drives the accelerated expansion of the Universe, the choice $f^2(R)=\mu R^m$, with $\mu>0$, was made.
Since the stability of the chameleon solution in the Solar System found in Ref. \cite{MBMGDeA,MBMDeA} requires $\mu<0$, in the present paper
we investigate the behavior of the cosmological tracking solution for $\mu<0$. With this choice we find that the addition of a cosmological constant
is still necessary in order that the tracking solution give rise to the Universe's accelerated expansion, nevertheless, the nonminimal coupling
between curvature and matter yields deviations from $\Lambda$CDM by reproducing the behavior of dynamical dark energy.

Analyzing this model in light of what recent data suggest for dynamical dark energy (DDE) arising from the DESI collaboration is particularly relevant, as in Ref. \cite{BarrosoVarela:2024ozs} the NMC gravity model with $\mu>0$ was numerically implemented and determined to be in better statistical agreement (compared to $\Lambda$CDM) 
with combinations of the DESI Data Release 1 (DR1) and Pantheon+/Dark Energy Survey (DES) supernovae distance moduli data, 
thus showing its promise in better explaining the currently debated signs of DDE. 

In the present paper we compare the tracking solution to
recent measurements of baryon acoustic oscillations (BAO) from the DESI collaboration in combination with the Pantheon+ and DES supernovae compilations.
We find constraints on the parameters of the NMC gravity model and we compare its fit properties to the Flat-$\Lambda$CDM model.
For suitable values of the allowed parameters the results indicate the presence of a dark energy evolving over time.
The model is further constrained by the results of the test of WEP based on LLR data. Though the set of allowed parameters is reduced, it turns out that
the LLR constraint does not exclude the values of parameters which give rise to an observable evolving dark energy.

This paper is organized as follows. In Section \ref{sec:NMC_Model}, we present the nonminimally coupled model under study. In Section \ref{sec:Field_Equations}, the field equations are presented and in Section \ref{sec:effective-potential} the properties of the effective potential that will play an important role in our discussion. In Section \ref{sec:Tracking_Solution}, the cosmological tracking solution used to test the model is introduced and its most salient properties are discussed. In Section \ref{sec:DarkEnergy_EqOfState}, the dark energy effective equation of state is discussed. In Section \ref{sec:Earth_Moon_Dynamics}, the main results about
the dynamics of the Earth-Moon system and the chameleon solution obtained in Ref. \cite{MBMDeA} are summarized leaving out the technical details.
In Section \ref{sec:Constraints_CosmologicalData}, we introduce the cosmological datasets used in constraining the model under study and combine these results with those of limits from LLR data. We finalize the paper by presenting our conclusions and prospects for future research in Section \ref{sec:Conclusions}.

%%%%%%%%%%%%%%%%%%%%%%%%%%%%%%%%%%%%%%%%%%%%%%%%%%%%%%%%%%%%%%%%%%%%%%%
\section{Nonminimally coupled gravity}\label{sec:NMC_Model}
%%%%%%%%%%%%%%%%%%%%%%%%%%%%%%%%%%%%%%%%%%%%%%%%%%%%%%%%%%%%%%%%%%%%%%%

We consider the action functional of NMC gravity theory of the form proposed in \cite{BBHL},
\begin{equation}\label{action-funct}
S = \int \left[\frac{1}{2}f^1(R) + [1 + f^2(R)] \LL_m \right]\sqrt{-g} \, d^4x,
\end{equation}
%\alpha\beta
where $f^i(R)$ (with $i=1,2$) are functions of the Ricci curvature scalar $R$, $\LL_m$ is the Lagrangian
density of matter, and $g$ is the metric determinant.
The Einstein-Hilbert action of GR is recovered by taking
\begin{equation}
f^1(R) = \frac{c^4}{8\pi G}R, \qquad f^2(R) = 0,
\end{equation}
where $G$ is Newton's gravitational constant. We work in the Jordan frame throughout this paper.

The first variation of the action with respect to the metric $g_{\alpha\beta}$ yields the field equations:
\begin{eqnarray}\label{field-eqs}
& &\left(f^1_R + 2f^2_R \LL_m \right) R_{\alpha\beta} - \frac{1}{2} f^1 g_{\alpha\beta}  \\
& &= \left(\nabla_\alpha \nabla_\beta -g_{\alpha\beta} \square \right) \left(f^1_R + 2f^2_R \LL_m \right)
+ \left(1 + f^2 \right) T_{\alpha\beta}, \nonumber
\end{eqnarray}
where $f^i_R \equiv df^i\slash dR$. The trace of the field equations is given by
\begin{eqnarray}\label{trace}
& &\left( f^1_R + 2f^2_R \LL_m \right) R - 2f^1 + 3\square f^1_R +
6\square\left( f^2_R \LL_m \right)  \nonumber\\
& &= \left( 1 + f^2 \right) T,
\end{eqnarray}
where $T$ is the trace of the energy-momentum tensor $T_{\alpha\beta}$.

By applying the Bianchi identities to Eq. (\ref{field-eqs}), it follows
\begin{equation}\label{covar-div-1}
\nabla_\alpha T^{\alpha\beta} = \frac{f^2_R }{ 1 + f^2} ( g^{\alpha\beta} \LL_m - T^{\alpha\beta} ) \nabla_\alpha R,
\end{equation}
so that, differently from GR and $f(R)$ gravity theory, in NMC gravity the energy-momentum tensor of matter is not covariantly conserved \cite{multiscalar,Sotiriou1}.

%%%%%%%%%%%%%%%%%%%%%%%%%%%%%%%%%%%%%%%%%%%%%%%%%%%%%%%%%%%%%%%%%%%%%%%%%%%%%
\subsection{Metric and energy-momentum tensors}\label{sec:metric-stress-tensor}
%%%%%%%%%%%%%%%%%%%%%%%%%%%%%%%%%%%%%%%%%%%%%%%%%%%%%%%%%%%%%%%%%%%%%%%%%%%%%

We use the following notation for indices of tensors:
Greek letters denote space-time indices ranging from 0 to 3,
whereas Latin letters denote spatial indices ranging from 1 to 3,
cartesian three-vectors are indicated by boldface type.
The signature of the metric tensor is $(-,+,+,+)$.

We consider a homogeneous, isotropic and spatially-flat universe, described by the Friedmann-Lema\^{i}tre-Robertson-Walker metric,
\begin{equation}\label{FLRW-metric}
ds^2 = -c^2 dt^2 + a^2(t) d\x^2,
\end{equation}
where $a$ is the cosmological scale factor, with $a(t_0)=1$, $t_0$ being the present time. For this metric
\begin{equation}\label{Ricci-curvature}
R = \frac{6}{c^2}\left(\frac{dH}{dt}+2H^2\right),
\end{equation}
where $H=(1\slash a)da\slash dt$ is the Hubble parameter. We consider a matter-dominated energy-momentum tensor,
\begin{equation}
T_{\alpha\beta} = \rho c^2 u_\alpha u_\beta,
\end{equation}
where $\rho$ is the average density. In comoving coordinates the four-velocity $u^\beta$ of cosmological matter is $u^\beta=(u^0,0,0,0)$, the normalization condition $u^\beta u_\beta = -1$
translates into $\left(u^0\right)^2=1$, so that the components of the energy-momentum tensor become $T_{00}=\rho c^2$ and all other components vanish. The trace of the tensor is $T=-\rho c^2$.

In the present paper we use $\LL_m = -\rho c^2$ for the Lagrangian density of matter \cite{BLP}.

Now we compute the zeroth component of Eq. (\ref{covar-div-1}) which gives the covariant derivative of the energy-momentum tensor:
\begin{equation}
\nabla_\alpha T^{\alpha 0} = \frac{f^2_R }{ 1 + f^2} ( g^{\alpha 0} \LL_m - T^{\alpha 0} ) \nabla_\alpha R.
\end{equation}
Using the definition of the energy-momentum tensor and of the matter Lagrangian we find
\begin{equation}
( g^{\alpha 0} \LL_m - T^{\alpha 0} ) \nabla_\alpha R = 0,
\end{equation}
from which it follows $\nabla_\alpha T^{\alpha 0}=0$ as in GR and $f(R)$ gravity theory, so that the usual equation for density evolution is obtained:
\begin{equation}\label{density-deriv}
\frac{d\rho}{dt} + 3H\rho = 0,
\end{equation}
with solution $\rho(t) = \rho_0 [a(t)]^{-3}$, where $\rho_0$ is the average density at the present epoch.

The recovery of the usual conservation law of GR is a consequence of the choice $\LL_m = -\rho c^2$ for the matter Lagrangian density. In Ref. \cite{Schutz:1970my}, Schutz obtained the relativistic perfect fluid Lagrangian density by utilizing a velocity potential approach in an Eulerian picture, with the four-velocity being expressed in terms of six velocity potentials. This procedure leads to the perfect fluid Lagrangian density $\mathcal{L}_m=p$. This differs from Brown’s approach \cite{Brown1993}, where the density is a derived quantity from the pressure. In the calculation done by Schutz, one finds a derivation that obtains only $\mathcal{L}_m=p$ and not $\mathcal{L}_m=-\rho c^2$.
    
The recovery of the usual conservation laws is unique to the choice $\mathcal{L}_m=-\rho c^2$, and thus if one adheres strictly to Schutz's formalism then the conservation law is necessarily modified. 
However, in the case of the late-time matter-dominated Universe, this implies that $\rho_m\propto a^{-3}(1+f^2)^{-1}$, such that the NMC effect exactly cancels out in the modified Friedmann equation term $(1+f^2)T_{\mu\nu}$. All other terms are dependent on $f^2_R\mathcal{L}_m$, which during matter domination is significantly suppressed, leading to a practically negligible contribution to the cosmological background dynamics. This means that in the case of $\mathcal{L}_m=p$ one is hopeless to find interesting dynamical behavior emerging on top of the cosmological constant, thus rendering any chance of fitting the DESI dynamical dark energy signatures quite impractical.
%%%%%%%%%%%%%%%%%%%%%%%%%%%%%%%%%%%%%%%%%%%%%%%%%%%%%%%%%%%%%%%%%%%%%%%%%%%%%
\subsection{\texorpdfstring{Choice of functions $f^1(R)$ and $f^2(R)$}{Choice of functions f1(R) and f2(R)}}
%%%%%%%%%%%%%%%%%%%%%%%%%%%%%%%%%%%%%%%%%%%%%%%%%%%%%%%%%%%%%%%%%%%%%%%%%%%%%

We choose the following functions:
\begin{equation}\label{f1-f2-specific}
f^1(R) = \frac{c^4}{8\pi G} (R-2\Lambda), \qquad f^2(R) = \mu R^m,
\end{equation}
where $m<0$ and $\mu<0$ are real numbers that have to be considered as parameters of the NMC model of gravity.

A function $f^2(R)$ with a negative exponent $m$ has been used in Ref. \cite{drkmattgal} to model the rotation curves of galaxies,
and in Ref. \cite{curraccel} to model the current accelerated expansion of the Universe. The negative sign of the weight $\mu$ is important to ensure
the existence and stability of the chameleon solution in the Solar System \cite{MBMGDeA,MBMDeA}.

The cosmological constant $\Lambda>0$ in Eq. (\ref{f1-f2-specific}) drives
the accelerated expansion of the Universe, while the nonminimal coupling between curvature and matter will give rise to deviations from $\Lambda$CDM.

%%%%%%%%%%%%%%%%%%%%%%%%%%%%%%%%%%%%%%%%%%%%%%%%%%%%%%%%%%%%%%%%%%%%%%%
\section{Field equations}\label{sec:Field_Equations}
%%%%%%%%%%%%%%%%%%%%%%%%%%%%%%%%%%%%%%%%%%%%%%%%%%%%%%%%%%%%%%%%%%%%%%%

In order to look for a solution of the field equations it is useful to introduce the following scalar field $\eta$, as in Refs. \cite{MBMGDeA,MBMDeA},
\begin{equation}\label{eta-definition}
\eta = f^1_R +2f^2_R \, \LL_m,
\end{equation}
so that the field equations Eqs. (\ref{field-eqs}) become
\begin{equation}
\eta R_{\alpha\beta} - \frac{1}{2} f^1 g_{\alpha\beta} = \left(\nabla_\alpha \nabla_\beta -g_{\alpha\beta} \square \right) \eta + \left(1 + f^2 \right) T_{\alpha\beta}.
\end{equation}
With the choice (\ref{f1-f2-specific}) of functions $f^1,f^2$ and $\LL_m = -\rho c^2$, the scalar field $\eta$
is a function of curvature $R$ also explicitly depending on cosmic time $t$ through mass density:
\begin{equation}\label{eta-expression}
\eta = \eta(t,R) = \frac{c^4}{8\pi G} -2 m\mu R^{m-1}\rho(t)c^2.
\end{equation}
Then, using the identities
\begin{eqnarray}
\nabla_0\nabla_0 \eta &=& \frac{1}{c^2}\frac{d^2\eta}{dt^2}, \\
\square\eta &=& -\frac{1}{c^2}\left(\frac{d^2\eta}{dt^2} + 3H\frac{d\eta}{dt}\right), \label{Dalemb-eta}
\end{eqnarray}
the time-time component of the field equations (\ref{field-eqs}) becomes the Friedmann equation modified in NMC gravity:
\begin{equation}\label{mod-Friedmann}
\frac{3}{c^2}H\frac{d\eta}{dt} - \frac{3}{c^2}\left(\frac{dH}{dt} + H^2\right)\eta + \frac{f^1}{2} = \left(1+f^2\right)\rho c^2.
\end{equation}
If $\mu=0$, this equation reduces to the Friedmann equation for the flat $\Lambda$CDM model.
The trace (\ref{trace}) of the field equations becomes
 \begin{equation}
 3\square\eta = 2 f^1 - \eta R - \left(1 + f^2\right)\rho c^2,
 \end{equation}
 which, by means of the introduction of the effective potential $V_{\rm eff}=V_{\rm eff}(\eta,\rho)$ defined by
 \begin{equation}\label{eff-potential}
 \frac{\partial V_{\rm eff}}{\partial\eta} = \frac{1}{3}\left[ 2 f^1 - \eta R - \left(1 + f^2\right)\rho c^2 \right],
 \end{equation}
 can be written in the form
 \begin{equation}
 \square\eta = \frac{\partial V_{\rm eff}}{\partial\eta}.
 \end{equation}
 The dependence of $V_{\rm eff}$ on $\eta$ results by expressing curvature as a function $R=\omega(\eta,\rho)$ obtained by solving Eq. (\ref{eta-expression}) with respect to $R$:
  \begin{equation}\label{curv-omega}
  R = \omega(\eta,\rho) = \left( \frac{c^4}{8\pi G} - \eta \right)^{1\slash(m-1)} \left( 2m\mu\rho c^2 \right)^{1\slash(1-m)}.
  \end{equation}
 Eventually, by using the identity (\ref{Dalemb-eta}), the trace equation becomes
 \begin{equation}\label{trace-eta}
 \frac{d^2\eta}{dt^2} + 3H\frac{d\eta}{dt} = -c^2\frac{\partial V_{\rm eff}}{\partial\eta}.
 \end{equation}
 We look for a solution of the modified Friedmann equation (\ref{mod-Friedmann}) and of the trace equation (\ref{trace-eta}),
 such that the scalar field $\eta$ follows the minimum of the effective potential during the Universe's expansion starting at the beginning of the
 matter dominated epoch. Hence, the next section is devoted to the study of the properties of the effective potential.
 
 %%%%%%%%%%%%%%%%%%%%%%%%%%%%%%%%%%%%%%%%%%%%%%%%%%%%%%%%%%%
\section{Properties of the effective potential}\label{sec:effective-potential}
%%%%%%%%%%%%%%%%%%%%%%%%%%%%%%%%%%%%%%%%%%%%%%%%%%%%%%%%%%%

We begin by determining the critical points of the effective potential for a given density $\rho$, which are the values of $\eta$ that satisfy the condition $\partial V_{\rm eff}\slash\partial\eta=0$.

\begin{figure}[t!]
    \centering
    \includegraphics[width=\linewidth]{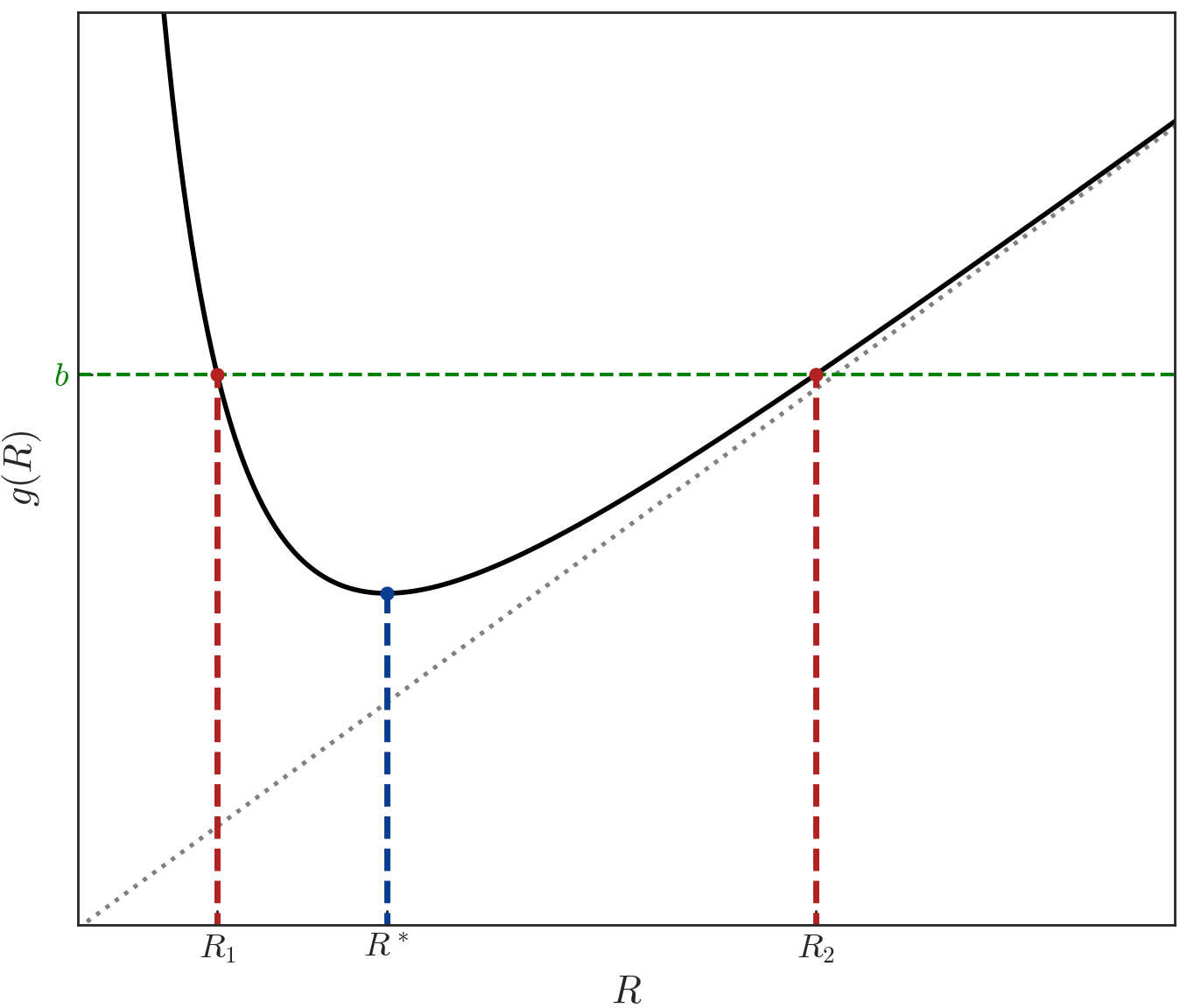}
    \caption{The properties of the critical points of the effective potential are shown through the auxiliary function $g(R)$ introduced in Eq. \eqref{g(R)-definition}. The function obeys the expected low and high curvature limits. Here we show the case $b>g(R^*)$, such that there exist two critical points, with $R_1$ corresponding to a maximum and $R_2$ to a minimum of the effective potential.}
    \label{fig:g(R)}
\end{figure}

Then, using Eq. (\ref{eff-potential}), the definition (\ref{f1-f2-specific}) of functions $f^1,f^2$, and the expression (\ref{eta-expression}) of the scalar field $\eta$,
the critical points are determined by the equation
\begin{equation}\label{critic-points-eq}
R + \frac{8\pi G}{c^2}(2m-1)\mu\rho R^m = \frac{8\pi G}{c^2}\rho + 4\Lambda.
\end{equation}
Now we set
\begin{eqnarray}
g(R) &=& R + d R^m, \label{g(R)-definition}\\
d &=&  \frac{8\pi G}{c^2}(2m-1)\mu\rho, \label{a-coeff-definition}\\
b &=& \frac{8\pi G}{c^2}\rho + 4\Lambda, \label{b-coeff-definition}
\end{eqnarray}
so that the critical points are the solutions of the equation $g(R)=b$, with $R=\omega(\eta,\rho)$. We show the qualitative behavior of $g(R)$ in Fig. \ref{fig:g(R)}.

Since $\Lambda>0$ we have $d>0$ and $b>0$, so that from $R>0$ it follows $g(R)>R>0$, moreover we have
\begin{equation}
\lim_{R\rightarrow 0^+}g(R) = +\infty, \qquad \lim_{R\rightarrow+\infty}\frac{g(R)}{R} = 1,
\end{equation}
and $g(R)$ is strictly convex. Hence, $g(R)$ has a unique minimum point $R^\ast$ given by
\begin{equation}
R^\ast = (-m d)^{\frac{1}{1-m}},
\end{equation}
which, using the expression (\ref{eta-expression}) of $\eta$, corresponds to
\begin{equation}\label{eta-inflection}
\eta^\ast =\frac{c^4}{8\pi G}\,\frac{1+2m}{2m-1}.
\end{equation}
Then, for $b>g(R^\ast)$ there exist two critical points, for $b=g(R^\ast)$ there exists one critical point, and for $b<g(R^\ast)$ there exist no critical points. We illustrate the first of these cases in Fig. \ref{fig:g(R)}.

Let us now consider the case $b>g(R^\ast)$ and let us determine the nature of the two critical points $\eta_1,\eta_2$ which, using Eq. (\ref{curv-omega}), corresponds to values
$R_1,R_2$ of curvature that satisfy $R_1<R^\ast<R_2$. Since, for a fixed value $\rho$ of density, $\eta$ is a monotone increasing function of $R$, we have
\begin{equation}\label{eta-ineq-chain}
\eta_1 < \eta^\ast < \eta_2.
\end{equation}
Using now Eq. (\ref{eff-potential}) and the expression (\ref{curv-omega}) of the function $R=\omega(\eta,\rho)$, we find
\begin{eqnarray}\label{V_eff-deriv2}
\frac{\partial^2 V_{\rm eff}}{\partial\eta^2} &=& [\omega(\eta,\rho)]^{2-m}\frac{1}{12m(1-m)\mu\rho c^2} \nonumber\\
&\times& \left[ \frac{c^4}{8\pi G}(1+2m) + (1-2m)\eta \right].
\end{eqnarray}
For fixed $\rho$, this second derivative vanishes at $\eta^\ast$, the effective potential is concave for $\eta<\eta^\ast$, and it is convex for $\eta>\eta^\ast$.
Hence, $\eta_1$ is a maximum point of $V_{\rm eff}$ and $\eta_2$ is a minimum point.

If $b=g(R^\ast)$, then $\eta^\ast$ is the unique critical point and it is a horizontal inflection point of $V_{\rm eff}$. If $b<g(R^\ast)$, then there are no critical points and the effective potential is
monotone decreasing. Moreover, note that the inflection point $\eta^\ast$ does not depend on density $\rho$.

\begin{figure}[t!]
    \centering
    \includegraphics[width=\linewidth]{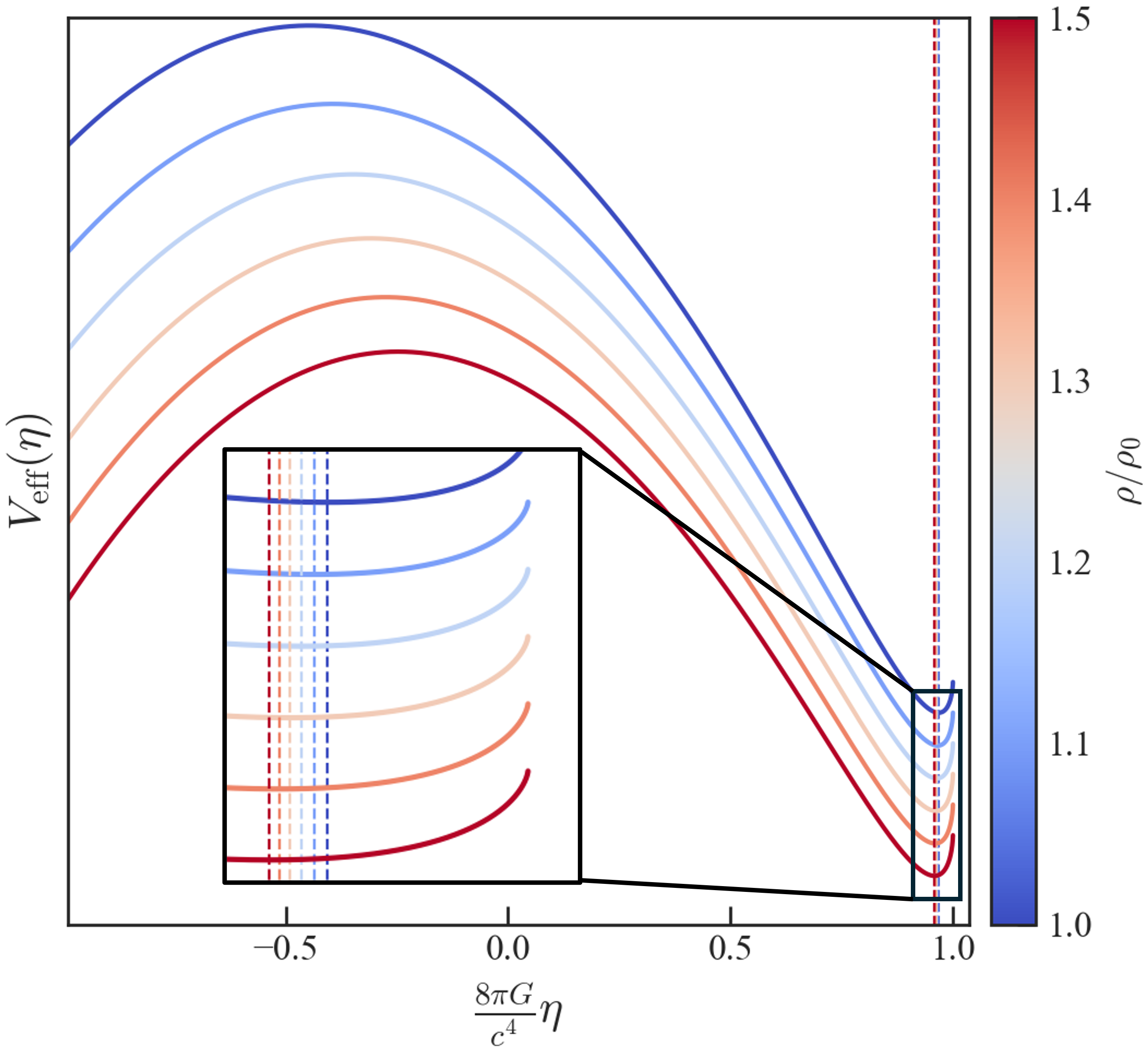}
    \caption{Effective potential of the scalar field $\eta$ for various densities of non-relativistic matter $\rho$ (or equivalent redshifts) in the $m=-3$ NMC model with $\mu^{1/|m|}=-4R_0$, $R_0$ being the curvature due to matter at present time. The cosmological constant is taken to be such that $\Omega_\Lambda=0.7$ and the potentials are shown in arbitrary units to focus on their relative magnitude. The location of the minimum of each potential is shown by a dashed line of the corresponding color.}
    \label{fig:VEff_plot}
\end{figure}

The full expression of the effective potential is obtained by integrating Eq. (\ref{eff-potential}) with respect to $\eta$ and expressing curvature by means of the function $R=\omega(\eta,\rho)$ given
in Eq. (\ref{curv-omega}):
\begin{eqnarray}\label{Veff-complete}
V_{\rm eff}(\eta,\rho) &=& \frac{1}{3}\left[(1-m)\mu\rho c^2(2m\mu\rho c^2)^{\frac{m}{1-m}}\Big(\frac{c^4}{8\pi G} - \eta\Big)^{\frac{m}{m-1}}\right. \nonumber\\
&\times& \left.\Big(\frac{3c^4}{8\pi G}-\eta\Big) - \rho c^2\eta - \frac{c^4}{2\pi G}\Lambda\eta\right] +\mathcal{C}(\rho),
\end{eqnarray}
where $\mathcal{C}(\rho)$ is an arbitrary additive function of density which does not change the equation Eq. (\ref{trace-eta}) for $\eta$. This potential is shown in Fig. \ref{fig:VEff_plot}. As expected, we see that the effective potential's minimum is shifted closer to $\eta=c^4/(8\pi G)$ as the non-relativistic matter density $\rho$ evolves, with potential's shape ensuring that the model does not stray significantly from GR, i.e. $8\pi G\eta/c^4=1$.

%%%%%%%%%%%%%%%%%%%%%%%%%%%%%%%%%%%%%%%%%%%%%%%%%%%%%%%%%%%%%%%%%
\subsection{Existence of the minimum for any density}\label{sec:minimum-any-density}
%%%%%%%%%%%%%%%%%%%%%%%%%%%%%%%%%%%%%%%%%%%%%%%%%%%%%%%%%%%%%%%%%

In this subsection we show that the presence of the additional cosmological term $\Lambda>0$ in the expression (\ref{f1-f2-specific}) of function $f^1(R)$ guarantees,
for suitable values of parameters $m$ and $\mu$,
the existence of the minimum of $V_{\rm eff}(\eta,\rho)$ for any density $\rho$ from the beginning of the matter dominated epoch to the whole future.

We have
\begin{eqnarray}
g(R^\ast) &=& \frac{8\pi G}{c^2}(1-m)(2m-1)\mu\rho \nonumber\\
&\times& \left[ -\frac{8\pi G}{c^2} m(2m-1)\mu\rho \right]^{\frac{m}{1-m}},
\end{eqnarray}
from which it follows
\begin{equation}
\lim_{\rho\rightarrow 0} g(R^\ast) = 0, \qquad \lim_{\rho\rightarrow 0}\frac{\rho}{g(R^\ast)} = 0,
\end{equation}
so that, if $\Lambda=0$, then for $\rho$ small enough during the Universe's expansion it follows $b<g(R^\ast)$ and there are no critical points. For $\Lambda>0$ the inequality $b>g(R^\ast)$
is equivalent to the following upper bound on $\vert\mu\vert$:
\begin{equation}\label{ineq-critic-points}
\vert\mu\vert < C(m) u(\rho),
\end{equation}
with
\begin{eqnarray}
u(\rho) &=& \rho^{-m}\left( 1 + \frac{c^2\Lambda}{2\pi G}\,\frac{1}{\rho} \right)^{1-m}, \label{u(rho)-function}\\
C(m) &=& \left(\frac{8\pi G}{c^2}\right)^{-m}\,\frac{1}{\vert 2m-1\vert}\left(\frac{1}{1-m}\right)^{1-m}\vert m\vert^{\vert m\vert}. \nonumber
\end{eqnarray}
The function $u(\rho)$ has a unique minimum point at density
\begin{equation}\label{minimum-density}
\rho_{\rm min} = \frac{1}{\vert m\vert}\,\frac{c^2\Lambda}{2\pi G},
\end{equation}
with value
\begin{equation}
u(\rho_{\rm min}) = \left(  \frac{1}{\vert m\vert}\,\frac{c^2\Lambda}{2\pi G} \right)^{-m}\left(1-m\right)^{1-m}.
\end{equation}
Hence, if $\vert\mu\vert < C(m) u(\rho_{\rm min})$, then inequality (\ref{ineq-critic-points}), consequently also inequality $b>g(R^\ast)$, is satisfied for any value of density $\rho$.
Substituting $\rho_{\rm min}$ in the inequality (\ref{ineq-critic-points}) we get the upper bound
\begin{equation}\label{mu-bound-any-density}
\vert\mu\vert < \frac{1}{\vert 2m-1\vert}(4\Lambda)^{\vert m\vert},
\end{equation}
which ensures the existence of the minimum and the maximum of $V_{\rm eff}(\eta,\rho)$ from the beginning of the matter dominated epoch to the whole future.
Note that $\Lambda>0$ is necessary to guarantee the existence of a positive upper bound.

Now we set $\eta_{\rm max}=\eta_1$ and $\eta_{\rm min}=\eta_2$, and we study the behavior of the two critical points as density $\rho$ tends to zero in the course of the Universe's expansion.
Using Eqs. (\ref{g(R)-definition}), (\ref{a-coeff-definition}) and (\ref{b-coeff-definition}), taking into account that $g(R_2)=b$, 
as $\rho\rightarrow 0$ we have $b\rightarrow 4\Lambda$ and $g(R_2)\rightarrow R_2$, from which, using expression (\ref{eta-expression}) for $\eta$, it follows
\begin{equation}
\lim_{\rho\rightarrow 0} R_2 = 4\Lambda, \qquad \lim_{\rho\rightarrow 0} \eta_{\rm min} = \frac{c^4}{8\pi G}.
\end{equation}
Note that the function $R=\omega(\eta,\rho)$ given in Eq. (\ref{curv-omega}) is well defined for any $m<0$ if $\eta<c^4\slash(8\pi G)$, so that as $\rho$ tends to zero
$\eta_{\rm min}$ converges to the supremum of $\eta$. 

Moreover, the value $R_2$ of curvature corresponding to $\eta_{\rm min}$ converges in the future to the curvature of De Sitter Universe.

We now consider the behavior of $\eta_{\rm max}$. Since in this case $g(R_1)=b$, using (\ref{a-coeff-definition}), as $\rho\rightarrow 0$ we have $R^\ast\rightarrow 0$, from which,
being $0<R_1<R^\ast$, it also follows $R_1\rightarrow 0$. Arguing as before we have $g(R_1)\rightarrow 4\Lambda$, so that, using Eqs. (\ref{g(R)-definition}) and (\ref{a-coeff-definition}),
we find
\begin{equation}
\lim_{\rho\rightarrow 0}\left(\rho R_1^m\right) = \frac{c^2}{2\pi G}\,\frac{\Lambda}{(2m-1)\mu},
\end{equation}
from which, using again expression (\ref{eta-expression}) for $\eta$, it follows
\begin{equation}
\lim_{\rm \rho\rightarrow 0} \eta_{\rm max} = -\infty.
\end{equation}
Taking the limit as $\rho$ tends to zero in Eq. (\ref{Veff-complete}), we find that the effective potential converges to an affine function:
\begin{equation}
\lim_{\rm \rho\rightarrow 0} V_{\rm eff}(\eta,\rho) = - \frac{c^4}{6\pi G}\Lambda\eta +\mathcal{C}(0),
\end{equation}
particularly, $\partial^2 V_{\rm eff}\slash\partial\eta^2\rightarrow 0$. However, using the minimum condition $g(R_2)=b$, setting $R_2=\omega(\eta_{\rm min},\rho)$
according to Eq. (\ref{curv-omega}), and using Eqs. (\ref{g(R)-definition}) and (\ref{V_eff-deriv2}), we find
\begin{equation}\label{second-deriv-diverges}
\lim_{\rm \rho\rightarrow 0} \frac{\partial^2 V_{\rm eff}}{\partial\eta^2}(\eta_{\rm min},\rho) = \lim_{\rm \rho\rightarrow 0}\frac{1}{\rho}  = +\infty.
\end{equation}
Hence, notwithstanding the effective potential converges to an affine function in $(-\infty,c^4\slash(8\pi G))$, 
the second derivative in the minimum diverges as the minimum point $\eta_{\rm min}$ approaches the supremum $c^4\slash(8\pi G)$ of $\eta$.

These asymptotic properties of the effective potential will be exploited in the subsequent sections in order to characterize the future Universe's expansion.

%%%%%%%%%%%%%%%%%%%%%%%%%%%%%%%%%%%%%%%%%%%%%%%%%%%%%%%%%%%%%%%
\section{Tracking solution}\label{sec:Tracking_Solution}
%%%%%%%%%%%%%%%%%%%%%%%%%%%%%%%%%%%%%%%%%%%%%%%%%%%%%%%%%%%%%%%

We follow now the approach of Ref. \cite{BvBDKW,FTBM}, and show the existence of a tracking solution consisting of the scalar field following the minimum of the effective potential, 
$\eta(t)=\eta_{\rm min}(t)$, which shifts as the Universe expands. The trace equation Eq. (\ref{trace-eta}) shows that the period $T$ of oscillations about the minimum is given by
\begin{equation}\label{oscill-period}
\left(\frac{2\pi}{T}\right)^2 = c^2 \frac{\partial^2 V_{\rm eff}}{\partial\eta^2}(\eta_{\rm min},\rho),
\end{equation}
and, according to Refs. \cite{BvBDKW,FTBM}, the condition for such a tracking solution to be valid is that the minimum satisfies
\begin{equation}\label{track-condition}
\left(\frac{TH}{2\pi}\right)^2 \ll 1,
\end{equation}
as the Universe expands. Moreover, the tracking solution is stable because if the scalar field $\eta$ is slightly perturbed away from the minimum $\eta_{\rm min}$,
it will oscillate and quickly settle back to the minimum \cite{BvBDKW}.

In this section we find bounds on parameters $m,\mu$ of the NMC gravity model which ensure the existence of the tracking solution
from the beginning of the matter dominated epoch to the whole future.

First we compute $H^2$ along the trajectory of the minimum $\eta_{\rm min}$. We begin by evaluating the term $H d\eta_{\rm min}\slash dt$ in the modified Friedmann equation Eq. (\ref{mod-Friedmann}).

Taking the derivative with respect to time of equation Eq. (\ref{critic-points-eq}) which determines the minimum, using the function $R_2=\omega(\eta_{\rm min},\rho)$, 
and using Eq. (\ref{density-deriv}) for the temporal derivative of density $\rho$,
we find
\begin{eqnarray}\label{H-eta-deriv1}
& &\left[ 1+(2m-1)m\mu\,\omega^{m-1}\frac{8\pi G}{c^2}\rho \right]\frac{d\omega}{dt} \nonumber\\
&=& -3H\,\frac{8\pi G}{c^2}\rho\left[ 1-(2m-1)\mu\,\omega^m \right].
\end{eqnarray}
Then, using expression (\ref{curv-omega}) of $\omega(\eta,\rho)$, and using again Eq. (\ref{density-deriv}), we have
\begin{equation}\label{H-eta-deriv2}
\frac{d\omega}{dt} = -\frac{\omega^{2-m}}{2m(m-1)\mu\rho c^2}\left( \frac{d\eta_{\rm min}}{dt}  - 6H\,m\mu\,\rho c^2\omega^{m-1}\right),
\end{equation}
where we consider $\omega=\omega(\eta_{\rm min},\rho)$.

Now, by substituting Eq. (\ref{H-eta-deriv2}) in Eq. (\ref{H-eta-deriv1}), and solving with respect to $d\eta_{\rm min}\slash dt$, we get
\begin{equation}\label{Fried-1}
H\frac{d\eta_{\rm min}}{dt} = 6H^2m\mu\,\omega^{m-2}\rho c^2\left( \omega + \frac{8\pi G}{c^2}\rho\,\frac{\widehat N}{\widehat D}\right),
\end{equation}
with
\begin{equation}\label{fraction-nHat-Dhat}
\frac{\widehat N}{\widehat D} = (m-1)\,\frac{1-(2m-1)\mu\,\omega^m}{1+(2m-1)m\mu\frac{8\pi G}{c^2}\rho\,\omega^{m-1}}.
\end{equation}
We now proceed with the computation of $H^2$. Using formula (\ref{Ricci-curvature}) for Ricci curvature scalar $R$, the following term of modified Friedmann Eq. (\ref{mod-Friedmann}) yields
\begin{equation}
-\frac{3}{c^2}\left( \frac{dH}{dt}+H^2 \right)\eta = \frac{3}{c^2}H^2\eta - \frac{R}{2}\eta,
\end{equation}
from which, using formula (\ref{eta-expression}) for $\eta$ and $R=\omega(\eta_{\rm min},\rho)$, it follows
\begin{eqnarray}\label{Fried-2}
& &-\frac{3}{c^2}\left( \frac{dH}{dt}+H^2 \right)\eta = \frac{3c^2}{8\pi G}\,H^2 - 6H^2 m\mu\,\rho\omega^{m-1} \nonumber\\
& &-\frac{\omega}{2}\left(\frac{c^4}{8\pi G} - 2m\mu\,\omega^{m-1}\rho c^2\right).
\end{eqnarray}
Moreover, using definition (\ref{f1-f2-specific}) of functions $f^1,f^2$, the other terms in the modified Friedmann equation Eq. (\ref{mod-Friedmann}) become
\begin{eqnarray}
\frac{f^1}{2} &=& \frac{c^4}{16\pi G}\,\omega - \frac{c^4}{8\pi G}\,\Lambda, \label{Fried-3}\\
\left(1+f^2\right)\rho c^2 &=& \rho c^2 + \mu\omega^m\rho c^2. \label{Fried-4}
\end{eqnarray}
Then, by substituting Eq. (\ref{Fried-1}) and Eqs. (\ref{Fried-2}-\ref{Fried-4}) in the modified Friedmann equation Eq. (\ref{mod-Friedmann}), and solving with respect to $H^2$, we obtain
\begin{equation}\label{Hubble-squared}
3H^2 = \frac{8\pi G\rho\left[ 1+(1-m)\mu\,\omega^m \right] + \Lambda c^2}{1 + 32\frac{\pi G}{c^2}\,\rho m\mu\,\omega^{m-2}\left(\omega + 12\frac{\pi G}{c^2}\,\rho\,\frac{\widehat N}{\widehat D}\right)}.
\end{equation}
Now we compute the period $T$ of oscillations about the minimum. Substituting formula (\ref{eta-expression}) for $\eta$ in the expression (\ref{V_eff-deriv2}) of
$\partial^2 V_{\rm eff}\slash\partial\eta^2$, and using formula (\ref{oscill-period}) for the period of oscillations, we obtain
\begin{equation}\label{period-computed}
\left(\frac{2\pi}{T}\right)^2 = \frac{\omega^{2-m}c^2\left[ 1 - \frac{8\pi G}{c^2}\rho m(1-2m)\mu\, \omega^{m-1} \right]}{48\frac{\pi G}{c^2}\rho m(1-m)\mu},
\end{equation}
where $\omega=\omega(\eta_{\rm min},\rho)$.
Note that the limit (\ref{second-deriv-diverges}) follows from this formula, so that, by definition (\ref{oscill-period}), the period $T$ of oscillations tends to zero as density $\rho$ tends to zero,
hence reinforcing condition (\ref{track-condition}) in the future expansion.

Eventually, using formulae (\ref{Hubble-squared}) and (\ref{period-computed}), we can evaluate the product $(TH\slash 2\pi)^2$ in condition (\ref{track-condition}). Nevertheless, it will be initially convenient
to transform the above formulae by introducing suitable auxiliary variables.

%%%%%%%%%%%%%%%%%%%%%%%%%%%%%%%%%%%%%%%%%%%%%%%%%%%%%%%%
\subsection{Auxiliary variables}
%%%%%%%%%%%%%%%%%%%%%%%%%%%%%%%%%%%%%%%%%%%%%%%%%%%%%%%%

In this section we introduce new variables which will allow us to achieve a more manageable expression of the product $(TH)^2$. In particular, that will be a low degree rational function of such variables
from which we will eventually obtain the product $(TH)^2$ as a function of the cosmological scale factor $a$ and the parameters of the NMC gravity model. 
Then we introduce the auxiliary variables $\xi$ and $Q$ defined by
\begin{equation}\label{xi-Q}
\xi = 1 - \frac{8\pi G}{c^4}\,\eta_{\rm min}, \qquad Q = \frac{8\pi G}{c^2}\,\frac{\rho}{\omega(\eta_{\rm min},\rho)}.
\end{equation}
Since $\eta<c^4\slash(8\pi G)$ we have $\xi>0$ and $Q>0$. We will find that condition (\ref{track-condition}) for the tracking solution to be valid is satisfied if the inequality $\xi\ll 1$ holds true.
Now we obtain some properties of the auxiliary variables that will be useful in the sequel.

For $\eta=\eta_{\rm min}$, formula (\ref{curv-omega}) which defines curvature as a function of $\eta$ and $\rho$, can be written in the form
\begin{equation}\label{Q-xi-simple}
2m\mu\,Q\omega^m = \xi,
\end{equation}
and equation (\ref{critic-points-eq}), which determines the minimum, can be written in the form
\begin{equation}\label{second-Q-formula}
Q = 1 - 4\,\frac{\Lambda}{\omega} + \frac{2m-1}{2m}\xi,
\end{equation}
from which, by using definition (\ref{xi-Q}) of $Q$ and solving with respect to $\omega$, we obtain the following relation which will have an interesting interpretation in the sequel:
\begin{equation}\label{omega-LCDM-dev}
\omega = \frac{\frac{8\pi G}{c^2}\rho + 4\Lambda}{1 + \frac{2m-1}{2m}\xi}.
\end{equation}
Then, by substituting formula (\ref{xi-Q}) for $Q$ and expression (\ref{omega-LCDM-dev}) for $\omega$ in relation (\ref{Q-xi-simple}), we obtain the following relation
between the parameters $m$ and $\mu$ of the NMC gravity model and the variable $\xi$:
\begin{equation}\label{m-mu-xi-relation}
\mu = \frac{1}{2m}\left(\frac{8\pi G}{c^2}\right)^{-m} u(\rho) y(\xi),
\end{equation}
where $u(\rho)$ is the function defined in Eq. (\ref{u(rho)-function}) and the function $y(\xi)$ is given by
\begin{equation}
y(\xi) = \frac{\xi}{\left(1 + \frac{2m-1}{2m}\,\xi\right)^{1-m}}.
\end{equation}
Since the function $u(\rho)$ has a unique minimum point at density $\rho_{\rm min}$ given by Eq. (\ref{minimum-density}), then for fixed values of parameters $m$ and $\mu$
the function $y(\xi)$ takes the maximum value at  density $\rho_{\rm min}$. Now, the function $y(x)$ is positive and monotone increasing for $x<x^\ast$, with $x^\ast$ maximum point given by
\begin{equation}
x^\ast = \frac{1}{1\slash 2 - m},
\end{equation}
then $y(x)$ is positive and monotone decreasing for $x>x^\ast$. Then, using inequalities (\ref{eta-ineq-chain}) and Eq. (\ref{eta-inflection}), we have
\begin{equation}
\xi = 1 - \frac{8\pi G}{c^4}\,\eta_{\rm min} < 1 - \frac{8\pi G}{c^4}\,\eta^\ast = x^\ast,
\end{equation}
so that, for fixed $m,\mu$, using relation (\ref{m-mu-xi-relation}), by decreasing density $\rho$ the variable $\xi$ increases reaching the maximum value $\xi_{\rm max}$ at density $\rho_{\rm min}$,
then $\xi$ decreases by further decreasing density below $\rho_{\rm min}$.

Using now $\rho=\rho_0 a^{-3}$ and formula (\ref{minimum-density}) for $\rho_{\rm min}$, we find the value $a_{\rm max}$ of the scale factor $a$ at which $\xi$ attains its maximum:
\begin{equation}\label{a-max}
a_{\rm max} = \left(\frac{\vert m \vert}{\chi}\right)^{1\slash 3}, \qquad \mbox{with} \qquad \chi = \frac{\Lambda c^2}{2\pi G\rho_0}.
\end{equation}
If we now set $R_0=(8\pi G\slash c^2)\rho_0$, which is the Ricci curvature scalar corresponding to the average density at the present epoch according to GR,
using $\rho=\rho_0 a^{-3}$, the relation Eq. (\ref{m-mu-xi-relation}) can be written in the form
\begin{equation}\label{relat-mu-xi}
\mu R_0^m  = \frac{a^{3m}}{2m}\left(1 + \chi a^3\right)^{1-m} y(\xi),
\end{equation}
which is now an equality between dimensionless quantities. Then, if we denote $\xi_0$ the value of $\xi$ at the present epoch, we have
\begin{equation}\label{relat-mu-xi0}
\mu R_0^m  = \frac{1}{2m}\left(1 + \chi\right)^{1-m} y(\xi_0),
\end{equation}
and using this last relation the pair of parameters $m,\mu$ of the NMC gravity model can be replaced with the pair $m,\xi_0$. Such a substitution turns out to be useful for suitable computations.

From formulae (\ref{relat-mu-xi}) and (\ref{relat-mu-xi0}) we find the relation between the variable $\xi$ and $\xi_0$ considered as a parameter:
\begin{equation}\label{relat-xi-xi0}
y(\xi) = a^{-3m}\left(\frac{1+\chi}{1+\chi a^3}\right)^{1-m}y(\xi_0).
\end{equation}
In the next section we will find that condition (\ref{track-condition}) for the tracking solution to be valid is satisfied, from the beginning of the matter dominated epoch to the whole future,
if the inequality $\xi\ll 1$ holds true over such a whole period of time. Since for $\xi\ll 1$ and $m$ of order unity we have $y(\xi)\approx \xi$, using relation Eq. (\ref{relat-mu-xi}),
we find the variable $\xi$ as a function of the scale factor $a$ which also depends on parameters $m$ and $\mu$:
\begin{equation}\label{relat-mu-xi-approx}
\xi \approx 2m\,a^{-3m}\frac{R_0^m}{\left(1+\chi a^3\right)^{1-m}}\,\mu,
\end{equation}
while, from relation Eq. (\ref{relat-xi-xi0}), we find $\xi$ as a function of $a$ which also depends on parameters $m$ and $\xi_0$:
\begin{equation}\label{relat-xi-xi0-approx}
\xi \approx a^{-3m}\left(\frac{1+\chi}{1+\chi a^3}\right)^{1-m}\xi_0.
\end{equation}
Using formula (\ref{a-max}) for $a_{\rm max}$, the maximum of $\xi$ is given by
\begin{equation}\label{eq:xi_max}
\xi_{\rm max} \approx \left(\frac{\chi}{\vert m\vert}\right)^m \left(\frac{1+\chi}{1-m}\right)^{1-m}\xi_0.
\end{equation}
Eventually, Eq. (\ref{relat-mu-xi0}) gives the relation between $\mu$ and $\xi_0$ for fixed $m$:
\begin{equation}\label{relat-mu-xi0-approx}
\mu R_0^m  \approx \frac{1}{2m}\left(1 + \chi\right)^{1-m} \xi_0.
\end{equation}
In the next section we will use the properties of the auxiliary variables $\xi$ and $Q$ in order to find a manageable form of the tracking condition (\ref{track-condition}).
\\

%%%%%%%%%%%%%%%%%%%%%%%%%%%%%%%%%%%%%%%%%%%%%%%%%%%%%%%%%%%%%%%%%%
\subsection{Existence of the tracking solution}
%%%%%%%%%%%%%%%%%%%%%%%%%%%%%%%%%%%%%%%%%%%%%%%%%%%%%%%%%%%%%%%%%%

An advantage of introducing the auxiliary variables will consist in the achievement of an expression of $(TH)^2$ which is a low degree rational function of the variable $\xi$, infinitesimal with respect to $\xi$,
and which also explicitly depends on the cosmological scale factor $a$. As a consequence, if $\xi\ll 1$, then the tracking condition (\ref{track-condition}) is satisfied.

We express the fraction $\widehat N\slash\widehat D$, given in Eq. (\ref{fraction-nHat-Dhat}) and that appears in formula (\ref{Hubble-squared}) for $H^2$, in terms of the auxiliary variables $\xi$ and $Q$:
\begin{equation}
\frac{\widehat N}{\widehat D} = (m-1)\frac{1-\frac{2m-1}{2m}\,\frac{\xi}{Q}}{1+\frac{1}{2}(2m-1)\xi}.
\end{equation}
Substituting this expression in formula (\ref{Hubble-squared}) for $H^2$, and using again the auxiliary variables, we obtain
\begin{equation}\label{H2-auxil-xi-Q}
H^2 = \frac{\frac{1}{12}\omega c^2\left(1+3Q+\frac{3-2m}{2m}\,\xi\right)}{1+2\xi\left[1+\frac{3}{2}(m-1)\frac{Q-\frac{2m-1}{2m}\,\xi}{1+\frac{1}{2}(2m-1)\xi}\right]}.
\end{equation}
\\
This formula for $H^2$ will be used in the sequel in order to obtain an effective equation of state for a dark energy model with the same expansion history.

The period (\ref{period-computed}) of oscillations is expressed by means of $\xi$ and takes the simple expression
\begin{equation}
\left(\frac{2\pi}{T}\right)^2 = \frac{\omega c^2}{3(1-m)\xi}\left[1-\frac{1}{2}(1-2m)\xi\right].
\end{equation}
Then the product $(TH\slash 2\pi)^2$ in the tracking condition (\ref{track-condition}) is given by
\begin{eqnarray}
& &\left(\frac{TH}{2\pi}\right)^2 = \frac{1}{4}(1-m)\xi \nonumber\\
&\times& \frac{1+3Q+\frac{3-2m}{2m}\xi}
{1+\frac{1}{2}(3+2m)\xi+\frac{2m-1}{2m}(3-m)\xi^2+3(m-1)\xi Q}. \nonumber\\
\end{eqnarray}
Since in the previous subsection we have found the expression of the variable $\xi$ as a function of the scale factor $a$ and the parameters of the NMC gravity model,
in order to obtain the final expression of $(TH)^2$ as a function of $a$ and the parameters, we now express the variable $Q$ as a function of $a$ and $\xi$, then we eliminate the
auxiliary variable $Q$ from the formula of $(TH)^2$.

Using formulae (\ref{second-Q-formula}) and (\ref{omega-LCDM-dev}) we have
\begin{equation}
Q = 1 - 4\Lambda\frac{1+\frac{2m-1}{2m}\,\xi}{\frac{8\pi G}{c^2}\,\rho+4\Lambda} + \frac{2m-1}{2m}\,\xi,
\end{equation}
from which, using $\rho=\rho_0 a^{-3}$ and the formula (\ref{a-max}) for $\chi$, we obtain
\begin{equation}\label{Q-funct-of-xi}
Q = \frac{1}{1+\chi a^3}\left(1 + \frac{2m-1}{2m}\,\xi\right).
\end{equation}
Substituting this formula for $Q$ in $(TH)^2$, we get the final expression:
\begin{widetext}
\begin{equation}\label{TH2-final-formula}
\left(\frac{TH}{2\pi}\right)^2 = \frac{1}{4}(1-m)\xi \,
\frac{1 + \frac{3-2m}{2m}\,\xi + \frac{3}{1+\chi a^3}\left(1+\frac{2m-1}{2m}\,\xi\right)}{1+\frac{1}{2}(3+2m)\xi+\frac{2m-1}{2m}(3-m)\xi^2+3\,\frac{m-1}{1+\chi a^3}\left(\xi+\frac{2m-1}{2m}\,\xi^2\right)}.
\end{equation}
\end{widetext}
This is a rational function of $\xi$ which also depends explicitly on the scale factor $a$. Moreover, $(TH)^2=\OO(\xi)$, so that the tracking condition (\ref{track-condition}) is satisfied
for $\xi\ll 1$. The first order approximation of this rational function with respect to $\xi$ yields
\begin{equation}
\left(\frac{TH}{2\pi}\right)^2 = \frac{1}{4}(1-m)\left(1+\frac{3}{1+\chi a^3}\right)\xi + \OO(\xi^2).
\end{equation}
Let $\eps$ be a positive number such that $\eps\ll 1$, then at first order in $\xi$ the tracking condition (\ref{track-condition}) can be written in the form
\begin{equation}
\frac{1}{4}(1+\vert m\vert)\left(1+\frac{3}{1+\chi a^3}\right)\xi < \eps.
\end{equation}
Using this inequality we can find an upper bound either on parameter $\vert\mu\vert$, or on parameter $\xi_0$, which ensures 
that condition (\ref{track-condition}) for the tracking solution to be valid, is satisfied from the beginning of the matter dominated epoch to the whole future.

Using Eq. (\ref{relat-mu-xi-approx}) that relates $\xi$ to $\mu$, and Eq. (\ref{relat-xi-xi0-approx}) that relates $\xi$ to $\xi_0$, we obtain, respectively,
\begin{eqnarray} 
\vert\mu\vert &<& \frac{2R_0^{-m}}{\vert m\vert(1+\vert m\vert)}\,v(a)\eps, \label{mu-bound-track} \label{eq:MaxMu_Tracking}\\
\xi_0 &<& \frac{4}{1+\vert m\vert}\left(1+\chi\right)^{m-1}v(a)\eps, \label{xi0-bound-track}
\end{eqnarray}

where the function $v(a)$ is given by
\begin{equation}
v(a) = \left(1+\frac{3}{1+\chi a^3}\right)^{-1}\left(1+\chi a^3\right)^{1-m}a^{3m}.
\end{equation}
For $a>0$ the function $v(a)$ has a unique minimum given by
\begin{equation}
\bar a = \left(\frac{h(m)}{\chi}\right)^{1\slash 3},
\end{equation}
with
\begin{equation}
h(m) = \frac{1}{2}\left[\vert m\vert -7 + \sqrt{(m+7)^2 + 16\vert m\vert}\right].
\end{equation}
Note that $\bar a$ is displaced with respect to $a_{\rm max}$ defined by Eq. (\ref{a-max}), so that $\xi(\bar a)<\xi_{\rm max}$ for a given $m$.

Then, by substituting $\bar a$ in inequalities (\ref{mu-bound-track}) and (\ref{xi0-bound-track}), we find upper bounds on $\vert\mu\vert$ and $\xi_0$ that ensure the existence of the tracking solution.
Such upper bounds define regions in the parameter planes with coordinates, $(\vert m \vert,\vert\mu\vert)$ and $(\vert m\vert,\xi_0)$, respectively, where the tracking condition (\ref{track-condition})
is satisfied. The upper bounds can be refined by resorting to the full formula (\ref{TH2-final-formula}) for $(TH\slash 2\pi)^2$.

These results have to be completed with the upper bounds on $\vert\mu\vert$ and $\xi_0$ that ensure the existence of the minimum of $V_{\rm eff}$ for any $a$ after the beginning of matter domination.
The bound on $\vert\mu\vert$ is given by inequality (\ref{mu-bound-any-density}) and the bound on $\xi_0$ follows from the relation (\ref{relat-mu-xi0-approx}):
\begin{eqnarray}
\vert\mu\vert &<& \frac{1}{\vert 2m-1\vert}(4\Lambda)^{\vert m\vert}, \label{eq:MaxMu_ExistenceOfMin} \\
\xi_0 &<&  \frac{2\vert m\vert}{\vert 2m-1\vert}\,\frac{1}{(1+\chi)^{1-m}}\left(\frac{4\Lambda}{R_0}\right)^{\vert m\vert} . \label{eq:MaxXi_ExistenceOfMin}
\end{eqnarray}
The intersection of bounds \eqref{mu-bound-track} and \eqref{xi0-bound-track} with \eqref{eq:MaxMu_ExistenceOfMin} and \eqref{eq:MaxXi_ExistenceOfMin} allows us to determine the allowed values of $\rho_0$ and $\Lambda$ for each chosen value of $|\mu|$. 
For each set of parameters $\{m,\mu,\rho_0,\Lambda\}$ we calculate $\chi$, $h(m)$, $\bar a$, $v(\bar a)$ and thus check if the conditions for the tracking solution and existence of minimum of $V_{\rm eff}$ are satisfied. In fact, since only the ratio $\rho_0/\Lambda$ is needed for this calculation, one can equivalently use the quantities $\Omega_m$ and $\Omega_\Lambda$, defined in terms of some unspecified present Hubble parameter $H_0$.
In the following we will make use of the quantity $|\mu|^{1/|m|}$ which defines a NMC curvature scale for a given $m$. This quantity has the same units as the Ricci scalar, {\em i.e.} $(\rm length)^{-2}$, which in cosmological contexts is typically chosen to be $\rm Mpc^{-2}$. To convert this to the same units as $H^2$, we may multiply it by $c^2=(3\times10^5 \ \rm km/s)^2$, therefore writing $|\mu|^{1/|m|}$ in $(\rm km/s/Mpc)^2$.

\begin{figure}[t!]
    \centering
    \includegraphics[width=\linewidth]{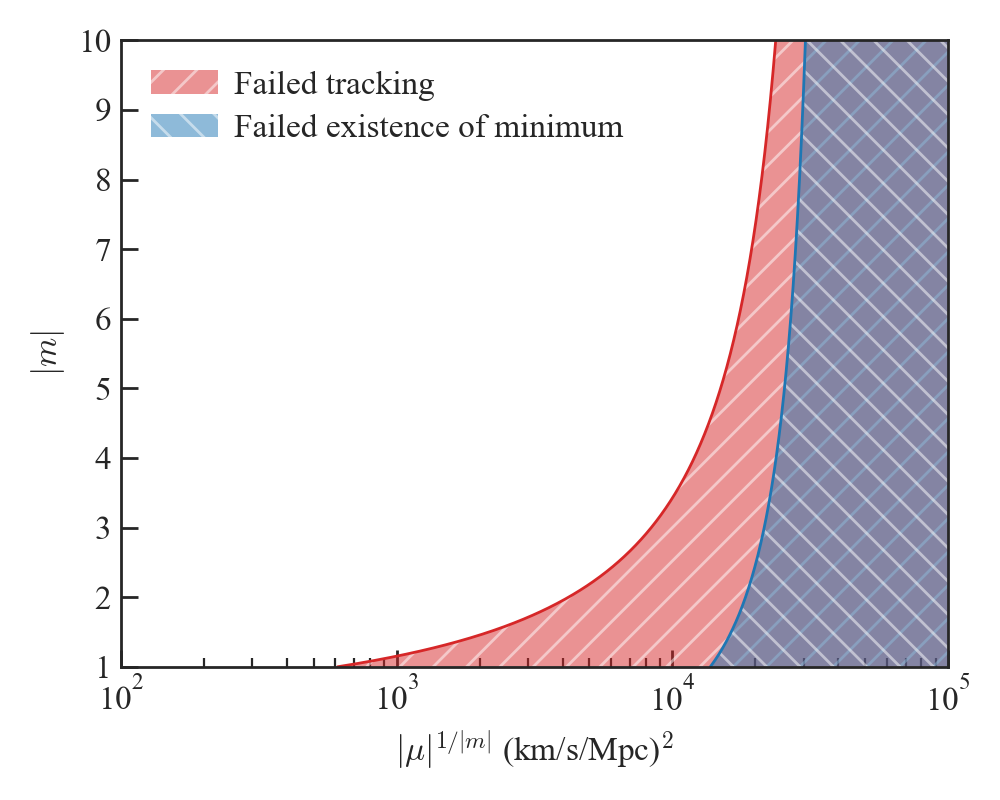}
    \caption{Allowed $\{|m|,|\mu|^{1/|m|}\}$ parameter space for typical values of matter ($\Omega_m=0.3$) and dark energy ($\Omega_\Lambda=0.7$) content in a Universe with $H_0=70 \ \rm km/s/Mpc$ and having fixed $\varepsilon=10^{-2}$. The tracking condition is always more stringent than the existence of minimum condition for the effective potential.}
    \label{fig:mu_m_Conditions}
\end{figure}

An example of the allowed $\{|m|,|\mu|^{1/|m|}\}$ parameter space is shown in Fig. \ref{fig:mu_m_Conditions}. Although in producing this figure we have chosen typical values of $\Omega_m=0.3$ and $\Omega_\Lambda=0.7$, we have verified that the qualitative nature of the results does not change for different values of $\Omega_i$. As seen in the figure, the tracking condition is always more stringent than the condition for the existence of the minimum of the effective potential for the scalar field $\eta$. Additionally, the magnitude of the coupling parameter $|\mu|^{1/|m|}$ is less constrained for larger exponents $|m|$, due to the scale factor $\bar a$ corresponding to the minimum of $v(a)$ shifting to higher values for larger $|m|$, thus increasing $v(\bar a)$ and the bound on $|\mu|^{1/|m|}$.

\begin{figure}[t!]
    \centering
    \includegraphics[width=\linewidth]{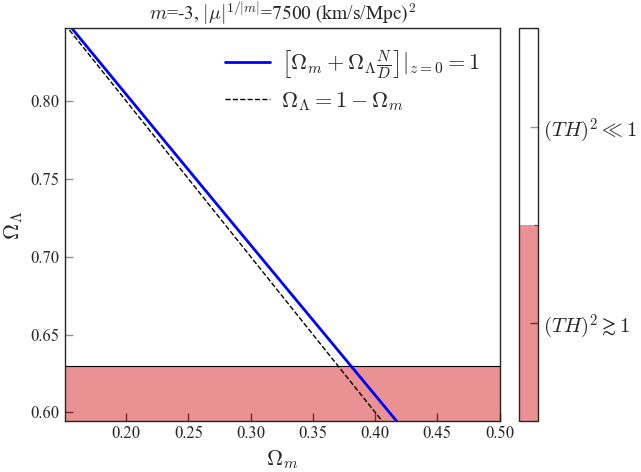}
    \caption{Allowed $\{\rho_0,\Lambda\}$ parameter space for a fixed exponent $m=-3$, NMC scale $|\mu|^{1/|m|}=7500 \ (\text{km/s/Mpc})^2$ and $\eps=10^{-2}$, with $H_0=70 \ \rm km/s/Mpc$ having been fixed for this plot. By increasing $|\mu|$, the disallowed (red) region is increased, as expected. The $H(0)=H_0$ constraint for the NMC model is shown in blue, showing its deviation from $\Lambda$CDM (dashed) as the Universe becomes more dynamical with an increasing dominance of matter content. }
    \label{fig:ParameterSpace}
\end{figure}

Another example of the allowed parameter space is shown in Fig. \ref{fig:ParameterSpace}. Although we have also enforced the existence of a minimum of $V_{\rm eff}$, note that we only show the tracking condition, as this always imposes stronger constraints on the parameter space (see Fig. \ref{fig:mu_m_Conditions}). As expected, the conditions require that the Universe be $\Lambda$-dominated, as the tracking solution depends on a slow evolution of the Universe, associated with the dominance of the more non-dynamical cosmological constant over the matter content, which is nonminimally coupled to the curvature, thus preferring regions of larger $\Omega_\Lambda$ and smaller $\Omega_m$. As we will discuss in the following section, the $\Omega_i$ parameters are constrained by the Friedmann equation, which reduces their allowed values to the blue line shown in Fig. \ref{fig:ParameterSpace}.

Now we note a property of the Ricci scalar curvature at the minimum of the effective potential, $R_2=\omega(\eta_{\rm min},\rho)$, given in Eq. (\ref{omega-LCDM-dev}) that we repeat here for convenience:
\begin{equation}
R_2 = \frac{\frac{8\pi G}{c^2}\rho + 4\Lambda}{1 + \frac{2m-1}{2m}\xi}.
\end{equation}
For $m$ of order unity and $\xi\ll 1$ curvature at the minimum is a perturbation of curvature of $\Lambda$CDM model. Moreover, Eq. (\ref{relat-xi-xi0-approx}) implies that as $a\rightarrow+\infty$,
$R_2$ converges to the curvature of De Sitter Universe (see also Section \ref{sec:minimum-any-density}).

Eq. (\ref{relat-xi-xi0-approx}) also shows that as $a$ decreases and it is small enough, 
$\xi$ decreases as $a^{3\vert m\vert}$, so that the initial condition for the scalar field $\eta$, at the beginning of the matter dominated epoch, corresponds to a curvature
that can be chosen close to the $\Lambda$CDM value.

If $\Lambda=0$ there exists a value of density $\rho^\ast$ such that the effective potential has no critical points for $\rho<\rho^\ast$. For $\xi\ll 1$, the tracking solution exists for $\rho>\rho^\ast$
and curvature at the minimum is a perturbation of curvature of GR theory, so that in this case the tracking solution is not able to reproduce the observed accelerated expansion of the Universe.

%%%%%%%%%%%%%%%%%%%%%%%%%%%%%%%%%%%%%%%%%%%%%%%%%%%%
\section{Dark energy effective equation of state}\label{sec:DarkEnergy_EqOfState}
%%%%%%%%%%%%%%%%%%%%%%%%%%%%%%%%%%%%%%%%%%%%%%%%%%%%

In this section we recast the impact of the nonminimal coupling between curvature and matter on the expansion history as an effective equation of state for a dark energy model with the same history.
We write the modified Friedmann equation (\ref{mod-Friedmann}) in terms of an effective dark energy density $\rho_X c^2$ as follows:
\begin{equation}
\frac{3}{8\pi G}\, H^2 = \rho + \rho_X,
\end{equation}
and the effective equation of state for the dark energy model is given by
\begin{equation}\label{de-equat-state}
w_X = -\frac{1}{3}\,\frac{d\ln\rho_X(a)}{d\ln a} - 1.
\end{equation}
This is the relevant equation of state that one would measure from the expansion history \cite{Hut-Shafer}.
We compute the equation of state along the trajectory of the minimum $\eta_{\rm min}$ of the effective potential.

First we write the expression of $\rho_X$ by resorting to the auxiliary variables $\xi$ and $Q$. We write Eq. (\ref{H2-auxil-xi-Q}) for $H^2$ in the form
\begin{equation}
\frac{3}{c^2}\, H^2 = \frac{1}{4}\,\frac{\omega A}{1+2\xi B},
\end{equation}
where
\begin{eqnarray}
A &=& 1 + 3Q + \frac{3-2m}{2m}\,\xi, \nonumber\\
B &=& 1 + \frac{3}{2}(m-1)\frac{Q-\frac{2m-1}{2m}\,\xi}{1+\frac{1}{2}(2m-1)\xi}.
\end{eqnarray}
Then, using definition (\ref{xi-Q}) of $Q$, we have
\begin{equation}
\rho_X = \frac{c^2}{8\pi G}\,\frac{1}{4}\,\frac{\omega A}{1+2\xi B} - \rho = \frac{\omega c^2}{8\pi G}\left(\frac{1}{4}\,\frac{A}{1+2\xi B}-Q\right),
\end{equation}
from which, using Eq. (\ref{second-Q-formula}), we obtain $\rho_X$ as a rational function of $\xi$ and $Q$:
\begin{equation}
\rho_X = \frac{c^2\Lambda}{8\pi G}\,\frac{N}{D},
\end{equation}
where
\begin{eqnarray}
N &=& A - 4Q(1+2\xi B), \nonumber\\
D &=& \left(1-Q+\frac{2m-1}{2m}\,\xi\right)(1+2\xi B).
\end{eqnarray}
If $\xi=0$ we have $N\slash D=1$, so that for $\xi$ small enough $\rho_X>0$ and $w_X$ in Eq. (\ref{de-equat-state}) is well defined. Indeed, we have checked the values of $\rho_X$ for parameters within the bounds of the tracking condition and found that, within this constraint, the dark energy density is always positive.

Now we observe that $N$ and $D$ can be expressed as functions of $a,m,\mu,\rho_0,\Omega_m\slash\Omega_\Lambda$,
so that we may write 
\begin{equation}\label{eq:HubbleEquation}
    H^2=H_0^2\left[\Omega_m(1+z)^3+\Omega_\Lambda\frac{N(z,m,\mu,\rho_0,\Omega_m\slash\Omega_\Lambda)}{D(z,m,\mu,\rho_0,\Omega_m\slash\Omega_\Lambda)}\right] \, ,
\end{equation}
where both $\Omega_i$ are defined as in $\Lambda$CDM and $z=1\slash a -1$ is the cosmological redshift. However, note that
we can no longer use $\Omega_m+\Omega_\Lambda=1$ as a constraint, which is replaced instead by
\begin{equation}
\Omega_m+\Omega_\Lambda\frac{N(0,m,\mu,\rho_0,\Omega_m\slash\Omega_\Lambda)}{D(0,m,\mu,\rho_0,\Omega_m\slash\Omega_\Lambda)}=1,
\end{equation}
which, being $N\slash D\approx 1$ for $\xi\ll1$, will be approximately equivalent to the former constraint, although not precisely identical. Both the $\Lambda$CDM and NMC constraints are shown in the $\Omega_m-\Omega_\Lambda$ plane in Fig. \ref{fig:ParameterSpace}. As expected, for a less dynamical $\Lambda$-dominated Universe ($\Omega_\Lambda\sim1$) we see that the NMC has little to no effect on the Friedmann equation and thus one recovers $\Lambda$CDM. Conversely, as we increase the fraction of matter in the present Universe we find that there can be non-negligible deviations from the standard Friedmann equation.

Using now definition Eq. (\ref{de-equat-state}), the dark energy effective equation of state becomes
\begin{equation}\label{equat-state-ratio-N-over-D}
w_X = - \frac{a}{3}\,\frac{N^\prime D - ND^\prime}{ND} - 1,
\end{equation}
where the prime denotes derivative with respect to $a$. In order to compute $w_X$ we require the derivatives of $\xi$ and $Q$ with respect to $a$.
Hence, taking the derivative with respect to $a$ of the identity Eq. (\ref{relat-mu-xi}) and solving with respect to $d\xi\slash da$, we obtain this derivative
as a rational function of $\xi$ multiplied by a rational function of $a$:
\begin{equation}
\frac{d\xi}{da} = -\left[ 3\frac{m}{a} + (1-m)\frac{3\chi a^2}{1+\chi a^3} \right]\frac{\left(2+\frac{2m-1}{m}\xi\right)\xi}{2+(2m-1)\xi}.
\end{equation}
Taking the derivative of $Q$, as given by Eq. (\ref{Q-funct-of-xi}), we have
\begin{equation}
\begin{aligned}
\frac{dQ}{da} = \frac{1}{1+\chi a^3}\left[ -\frac{3\chi a^2}{1+\chi a^3}\left(1+\frac{2m-1}{2m}\xi\right) \right.
\\ \left.+ \frac{2m-1}{2m}\frac{d\xi}{da} \right].
\end{aligned}
\end{equation}
Substituting these derivatives in formula (\ref{equat-state-ratio-N-over-D}), we obtain the following form for the equation of state:
\begin{equation}
w_X = \frac{N_{w_X}}{D_{w_X}} - 1,
\end{equation}
where $N_{w_X}$ and $D_{w_X}$ are polynomial functions of $\xi$ with a degree of 6 and coefficients that depend on the scale factor $a$ and parameter $m$.
Then the complete dependence of $w_X$ on $a$ and NMC gravity parameters results from solving Eq. (\ref{relat-mu-xi}) with respect to $\xi$ as a function of $a$ and $m,\mu$.
That is guaranteed by the invertibility of the function $y(\xi)$.
Furthermore, we have $w_X+1=\OO(\xi)$.

The complete result for the equation of state is reported in Appendix \ref{sec:Appendix_effectiveDarkEnergy}.
For $\xi\ll 1$, the first order approximation with respect to $\xi$ of the fraction $N_{w_X}\slash D_{w_X}$ yields
%
% \begin{widetext}
% \begin{eqnarray}\label{equat-state-first-order}
% w_X &=& \xi\left\{V^{(1)}_{0,1} + \frac{V^{(1)}_{1,1}}{1+\chi a^3} + \frac{V^{(1)}_{2,1}}{(1+\chi a^3)^2} + \frac{V^{(1)}_{3,1}}{(1+\chi a^3)^3}
% - \chi a^3 \left[ \frac{V^{(2)}_{1,1}}{1+\chi a^3} + \frac{V^{(2)}_{2,1}}{(1+\chi a^3)^2} +\frac{V^{(2)}_{3,1}}{(1+\chi a^3)^3} + \frac{V^{(2)}_{4,1}}{(1+\chi a^3)^4} \right]\right\} \nonumber\\
% &\times& \left[W_{0,0} + \frac{W_{1,0}}{1+\chi a^3} + \frac{W_{2,0}}{(1+\chi a^3)^2} + \frac{W_{3,0}}{(1+\chi a^3)^3} + \frac{W_{4,0}}{(1+\chi a^3)^4}\right]^{-1} -1 + \OO(\xi^2),
% \end{eqnarray}
% \end{widetext}
%
\begin{align}\label{equat-state-first-order}
    w_X=-1+\xi\frac{\sum_{k=0}^3 V_{k,1}(1+\chi a^3)^{k}}{\sum_{k=0}^3 W_{k,0}(1+\chi a^3)^{k}} \, ,
\end{align}
where the coefficients $V_{k,1}$ and $W_{k,0}$ are polynomial functions of parameter $m$. Formulae for these coefficients are reported in Appendix \ref{sec:Appendix_Coefficients}.
Then, substituting in Eq. (\ref{equat-state-first-order}) the variable $\xi$ with its approximations for $\xi\ll 1$ given in Eq. (\ref{relat-mu-xi-approx}) and Eq. (\ref{relat-xi-xi0-approx}),
we obtain $w_X$ as a function of $a,m,\mu$ and $a,m,\xi_0$, respectively. 

Using Eqs. (\ref{relat-mu-xi-approx}-\ref{relat-xi-xi0-approx}) we have that, as the scale factor $a$ tends to infinity,
$w_X+1$ tends to zero as $a^{-3}$, so that in the future we have convergence to De Sitter Universe.

Eventually, the resulting expression of the effective dark energy density $\rho_X c^2$ is reported in Appendices A and B.

%%%%%%%%%%%%%%%%%%%%%%%%%%%%%%%%%%%%%%%%%%%%%%%%%%%%
\section{Dynamics of the Earth-Moon system}\label{sec:Earth_Moon_Dynamics}
%%%%%%%%%%%%%%%%%%%%%%%%%%%%%%%%%%%%%%%%%%%%%%%%%%%%

In this section we summarize the main results achieved in Ref. \cite{MBMDeA} about the effects of the nonminimal coupling between curvature and matter on the dynamics of Earth and Moon
in the gravitational field of the Sun. The main effect is a violation of the weak equivalence principle.

The metric tensor which describes the spacetime in the Sun-Earth-Moon system is a perturbation of the flat FLRW metric (\ref{FLRW-metric}) at the present epoch $t_0$, when $a(t_0)=1$.
In the Newtonian gauge the metric tensor is given by
\begin{eqnarray} \label{metric}
ds^2= &-& \left[1-2\Phi(\x,t)+2\Psi(\x,t)\right]c^2dt^2 \nonumber\\
&+& \left[1+2\Phi(\x,t)\right] \delta_{ij}dx^idx^j,
\end{eqnarray}
where the potentials $\Phi$ and $\Psi$ are perturbations of the Minkowski metric of order $\OO(1/c^2)$, and time $t$ here refers to local motion.

The three astronomical bodies are modeled as spherically symmetric distributions of matter.
The Sun is considered as a perfect fluid in hydrostatic equilibrium, while the Earth and Moon are approximately described as continuous bodies
in a hydrostatic state of stress, {\em i.e.}, the normal stresses are equal to the pressure and shear stresses are neglected \cite{Turcotte}.
The components of the energy-momentum tensor, to the relevant order for our computations, for all the astronomical bodies
are given by (Ref. \cite{Wi}, Chapter 4.1):
\begin{eqnarray}
T^{00} &=& \rho c^2 + \OO\left(1\right), \label{T-00}\\
T^{0i} &=& \rho c v^i + \OO\left(\frac{1}{c}\right), \label{T-0i}\\
T^{ij} &=& \rho v^i v^j + p\delta_{ij} + \OO\left(\frac{1}{c^2}\right), \label{T-ij}
\end{eqnarray}
where matter is characterized by density $\rho$, velocity field $v^i$, and pressure $p$.
The trace of the energy-momentum tensor is
\begin{equation}
T = -\rho c^2 + \OO\left(1\right).
\end{equation}
The field equations (\ref{field-eqs}-\ref{trace}) have been approximated in Ref. \cite{MBMDeA}, for weak field and slow motion,
under the the following assumptions, that have been verified a posteriori:
\begin{equation}
\left\vert f^2(R) \right\vert \ll 1, \qquad \left\vert \frac{8\pi G}{c^4}\eta -1 \right\vert \ll 1,
\end{equation}
where the scalar field $\eta$ is still defined by Eq. (\ref{eta-definition}). Keeping terms of order $\OO(1/c^2)$, the field equations for the metric potentials $\Phi$ and $\Psi$ found in
Ref. \cite{MBMDeA} are given by
\begin{eqnarray}
\nabla^2\Phi &=& -\frac{4\pi G}{c^2}\rho + \frac{1}{6}\left( \frac{8\pi G}{c^2}\rho - R \right), \label{Phi-equation-approx}\\
\nabla^2\Psi &=& \frac{1}{3}\left(\frac{8\pi G}{c^2}\rho - R\right). \label{Psi-equation-approx}
\end{eqnarray}
The approximation of the trace equation (\ref{trace}) yields
\begin{equation}\label{trace-approx}
\nabla^2\eta = \frac{c^4}{24\pi G}\, R - \frac{1}{3}\rho c^2.
\end{equation}
By introducing a potential function $V=V(\eta,\rho)$ and an effective potential $V_{\rm eff}$ as in Refs. \cite{KW,HS},
\begin{equation}\label{potential-V-def}
\frac{\partial V}{\partial\eta}= \frac{c^4}{24\pi G}\,\omega(\eta,\rho), \qquad V_{\rm eff}=V-\frac{1}{3}\rho c^2\eta,
\end{equation}
where the function $\omega(\eta,\rho)$ is still given by formula (\ref{curv-omega}), the equation for the scalar field $\eta$ becomes
\begin{equation}\label{eta-equation-potential}
\nabla^2 \eta=\frac{\partial V_{\rm eff}}{\partial\eta}.
\end{equation}
The effective potential has an extremum which corresponds to the GR solution $R=\omega(\eta,\rho) = (8\pi G\slash c^2)\rho$,
and the requirement that such an extremum is a minimum, for $m<0$ yields the condition $\mu<0$ \cite{MBMGDeA,MBMDeA}.
This is an application of a general stability condition against Dolgov-Kawasaki instability in NMC gravity found in Refs. \cite{Faraoni,BeSeq}.

The effective potential defined in Eq. (\ref{potential-V-def}) is different from the effective potential studied in Sec. \ref{sec:effective-potential} for the cosmological field equations:
in the cosmological case $V_{\rm eff}$ depends on time through $\eta$ and cosmological density, while in the Sun-Earth-Moon system $V_{\rm eff}$ depends both on time and spatial
coordinates through $\eta$ and density of the moving bodies. In the latter case the properties of the effective potential have been studied in Ref. \cite{MBMDeA}.

At the minimum of the effective potential $V_{\rm eff}$ we set
\begin{equation}\label{lambda-definition}
\frac{\partial^2 V_{\rm eff}}{\partial\eta^2} = \frac{1}{\lambda^2} >0,
\end{equation}
where $\lambda=\lambda(\rho)>0$ has dimension of length and depends on mass density. We have \cite{MBMDeA}
\begin{equation}\label{lambda-expression}
\lambda^2 = 6 \mu m(1-m) \left(\frac{8\pi G}{c^2}\rho\right)^{m-1},
\end{equation}
and the function $\lambda(\rho)$ decreases as density $\rho$ increases.

%%%%%%%%%%%%%%%%%%%%%%%%%%%%%%%%%%%%%%%%%%%%%%%%%%%%
\subsection{Chameleon solution}
%%%%%%%%%%%%%%%%%%%%%%%%%%%%%%%%%%%%%%%%%%%%%%%%%%%%

The equation (\ref{eta-equation-potential}) for $\eta$ is typical of chameleon theories of gravity \cite{KW}, modified through the explicit dependence of the potential $V$ on density due to the nonminimal coupling.
The application of the chameleon mechanism to NMC gravity has been studied in Ref. \cite{MBMGDeA} for the gravitational field of the Sun.
In Ref. \cite{MBMDeA} Eq. (\ref{eta-equation-potential}) has been solved by extending to the Sun-Earth-Moon system the approach followed in Ref. \cite{MBMGDeA} for the one body problem.
Particularly, the nonlinear equation for $\eta$ has been approximated by means of different linear equations in different regions of space and the solutions of the linear equations
have then been matched across the boundaries of such regions by resorting to suitable boundary conditions \cite{MBMGDeA,KW,Kraiselburd,TaTsu}.

Following Ref. \cite{KW}, different approximations of the effective potential $V_{\rm eff}$ have been used in Ref. \cite{MBMDeA}. If the solution $\eta$ has to remain close to a minimum point $\bar\eta$
of $V_{\rm eff}$, then a quadratic approximation of the potential around the minimum has been used, so that the derivative of $V_{\rm eff}$ is approximately given by
\begin{equation}\label{quadr-approx-V_eff}
\frac{\partial V_{\rm eff}}{\partial\eta}(\eta,\rho) \approx \frac{\partial V_{\rm eff}}{\partial\eta}(\bar\eta,\rho) +
\frac{\partial^2 V_{\rm eff}}{\partial\eta^2}(\bar\eta,\rho)(\eta-\bar\eta).
\end{equation}
In order to satisfy the stringent bounds from Solar System experiments on modified gravity, a chameleon theory requires the solution $\eta$ to remain close
to the minimum point of the effective potential $V_{\rm eff}$ in most of the interior of massive astronomical bodies such as the Sun, Earth and Moon, so that GR
is approximately satisfied \cite{KW,HS}. More precisely, in each body $\eta$ has to be close to the minimum point of $V_{\rm eff}$ inside a critical radius, 
called the screening radius, that has to be determined. If the screening radius is close to the radius of the astronomical body for each body, then
the thin shell condition is satisfied and deviations from GR are screened \cite{KW}.

In Ref. \cite{MBMDeA} both the Earth and Moon have been modeled by means of layers of constant density, so that the approximation (\ref{quadr-approx-V_eff}) of the effective potential for
large density, inside the screening radii of Earth and Moon, yields
\begin{equation}\label{Yuk-approx-inside-rscreen}
\nabla^2\eta = \frac{\partial V_{\rm eff}}{\partial\eta}(\eta,\rho) \approx
\frac{1}{\lambda^2(\rho)}(\eta-\bar\eta),
\end{equation}
hence Eq. (\ref{eta-equation-potential}) for $\eta$ becomes of Yukawa type in each layer. 
Inside the screening spheres of the Earth and Moon it turns out that the difference between $\eta$ and $\bar\eta$ is exponentially suppressed in each layer due
to the smallness of $\lambda(\rho)$ for the large values of density, and curvature is close to the GR value $R\approx(8\pi G\slash c^2)\rho$ \cite{MBMDeA}.
For the Sun a continuous density profile has been adopted that requires a different approach discussed in \cite{MBMGDeA}.

Following Refs. \cite{KW,HS}, in Ref. \cite{MBMDeA} the solution $\eta$ was required to be close to the minimum point of $V_{\rm eff}$ also in the outskirts of the Solar System where $\rho$ approaches the
galactic mass density, so that at large distances from the Sun's center beyond the boundary of the heliosphere the quadratic approximation 
(\ref{quadr-approx-V_eff}) of the effective potential has been used again, in this case for low density. 
Since the function $\lambda(\rho)$ increases as density $\rho$ decreases it turns out that in this region $\lambda(\rho)$ is large
enough to turn Yukawa equation (\ref{Yuk-approx-inside-rscreen}) into Laplace equation
\begin{equation}
\nabla^2\eta = 0,
\end{equation}
with the boundary condition $\eta\approx\eta_g$ at large distance from the Sun, $\eta_g$ being
the minimum point of $V_{\rm eff}$ in the solar neighborhood of the Milky Way,
\begin{equation}\label{eta-g-minimizer}
\eta_g = \frac{c^4}{8\pi G} - 2m\mu\left(\frac{8\pi G}{c^2}\right)^{m-1}\rho_g^m c^2,
\end{equation}
obtained substituting in formula (\ref{eta-expression}) the corresponding values of the galactic density $\rho_g$ and GR curvature \cite{MBMDeA}.

Now we consider the thin shells of the Sun, Earth and Moon. For each of the astronomical bodies the thin shell is the spherical shell defined by $r_s<\vert\x-\x_b\vert<r_b$,
where $\x_b$ is the position vector of the center of the body, $r_s$ is the screening radius and $r_b$ is the radius of the body.
We call the sphere with center $\x_b$ and radius $r_s$ a screening sphere.

Following again \cite{KW}, in Ref. \cite{MBMDeA} inside the three thin shells the inequality $\partial V\slash\partial\eta \ll \rho c^2$ has been used, so that from Eq. (\ref{potential-V-def})
the derivative of the effective potential $V_{\rm eff}$ has been approximated by means of
\begin{equation}
\nabla^2\eta = \frac{\partial V_{\rm eff}}{\partial\eta} \approx - \frac{1}{3}\rho c^2,
\end{equation}
which corresponds to consider curvature $R$ much smaller than the GR value inside the thin shells.

In the interplanetary space between the bodies mass density is low and it turns out that the equation for $\eta$ is well approximated by Laplace equation $\nabla^2\eta=0$. 
Hence, outside of the screening spheres of the three bodies up to the outskirts of the Solar System, at each time instant $t$ the equation for $\eta$ is Poisson equation
with source different from zero in the thin shells and given by density $\rho$ in the shells. 
Both the domain of the Poisson equation and the source $\rho$ change with time as the bodies move in interplanetary space, nevertheless, at each $t$ a static problem has to be solved \cite{MBMDeA}.

In order to ensure the existence of $\nabla^2\eta$, matching conditions that impose the continuity of $\eta$ 
and its space derivatives are imposed across the screening spheres of the three bodies, moreover, $\eta$ has to approach $\eta_g$ at the outskirts of the Solar System.

Continuity of $\eta$ imposes a Dirichlet boundary condition on each screening sphere, and
the solution of the resulting Dirichlet boundary problem for Poisson equation in the thin shells and in interplanetary space has been computed in Ref. \cite{MBMDeA} by means of Green's function method.
In the case of a single sphere the Green's function could be obtained by using the method of images as
in electrostatics, however, since the Dirichlet boundary condition is given on three spheres, an extension of the method of images to a system of spheres \cite{MetzLa} has been used in
Ref. \cite{MBMDeA} in order to compute the Green's function. 
Eventually, continuity of the space derivatives of $\eta$ across the screening spheres has been used in Ref. \cite{MBMDeA} to determine
the screening radii of the three bodies which are among the unknowns of the problem.

The solution $\eta$ for the Sun-Earth-Moon system is the following:
\begin{equation}\label{eta-solut-Green}
\eta = \eta_S + \eta_E + \eta_M + \eta_g,
\end{equation}
where $\eta_S,\eta_E$ and $\eta_M$ are the contributions from the thin shells of Sun, Earth and Moon, respectively.
The contribution from the Earth is
\begin{equation}
\eta_E =  \frac{c^2}{12\pi}\,\frac{M_{\oplus,{\rm eff}}}{\vert\x-\x_E\vert} + \OO\left(\frac{R}{d}\right) \qquad\mbox{for } \vert\x-\x_E\vert \geq R_\oplus,
\end{equation}
where $\x_E$ and $R_\oplus$ are the position vector of the center of Earth and the radius of Earth, $M_{\oplus,{\rm eff}}$ is the effective mass of Earth, that is the mass of the thin shell,
\begin{equation}\label{Earth-effective-mass}
M_{\oplus,{\rm eff}} = 4\pi \int_{r_E}^{R_\oplus} \rho_E(r)r^2dr\,,
\end{equation}
$r_E$ and $\rho_E$ being the screening radius and density of Earth, respectively, and $\OO(R/d)$ denotes terms multiplied by a factor of type $R\slash d$, 
where $R$ is a radius and $d$ is a distance between the three astronomical bodies. The terms $\OO(R/d)$, reported in Ref. \cite{MBMDeA},
result from the application of the method of images for a system of spheres and they take into account the interactions between the three bodies.
The contributions of Sun and Moon to $\eta$ have analogous expressions.

The screening radii of the three bodies are determined by a system of integral equations \cite{MBMDeA}. The screening radius $r_E$ of Earth is the lower limit of integration in the equation
\begin{equation}\label{Earth-screening-radius}
\frac{c^2}{3}\int_{r_E}^{R_\oplus}\rho_E(r)r\,dr  =  \frac{c^4}{8\pi G} - \eta_g + \OO\left(\frac{R}{d}\right),
\end{equation}
with $\eta_g$ given by Eq. (\ref{eta-g-minimizer}), so that the screening radius depends on the NMC gravity parameters $m$ and $\mu$.
Analogous integral equations hold for the Sun and for the Moon.

We now give the solution for the metric potentials $\Phi$ and $\Psi$.
Using Eqs. (\ref{Phi-equation-approx}-\ref{Psi-equation-approx}) it follows that the potential $\Psi$ is related to the deviation from GR, then, in Ref. \cite{MBMDeA}
the following boundary conditions have been imposed in the solar neighborhood of the Galaxy at large enough distance from the Sun's center, where GR is approximately satisfied by the assumption that
$\eta$ is close to the minimum point of the effective potential $V_{\rm eff}$:
\begin{equation}\label{Phi-Psi-boundcond-infty}
\Phi(\x,t) \approx \frac{1}{c^2}U(\x,t), \qquad \Psi(\x,t) \approx 0,
\end{equation}
where $U$ is the Newtonian potential. Combining Eqs. (\ref{Psi-equation-approx}) and (\ref{trace-approx}) for $\Psi$ and $\eta$ we have
\begin{equation}
\nabla^2 \left( \Psi + \frac{8\pi G}{c^4}\, \eta \right) = 0,
\end{equation}
then, using boundary conditions (\ref{Phi-Psi-boundcond-infty}) for $\Psi$ and $\eta\approx\eta_g$,
an application of the maximum principle for harmonic functions yields the solution for the potential $\Psi$ found in Ref. \cite{MBMDeA}:
\begin{equation}\label{potential-Psi-sol}
\Psi = - \frac{8\pi G}{c^4}\left( \eta - \eta_g \right).
\end{equation}
The solution for $\Psi$ then follows immediately from the solution for $\eta$ given in Eq. (\ref{eta-solut-Green}).
Combining now equations (\ref{Phi-equation-approx}-\ref{Psi-equation-approx}) for $\Phi$ and $\Psi$ we have
\begin{equation}
\nabla^2\left( \Phi - \frac{U}{c^2} -\frac{1}{2}\, \Psi \right) = 0,
\end{equation}
from which, applying boundary conditions (\ref{Phi-Psi-boundcond-infty}) and using again the maximum principle for harmonic functions,
the solution for the potential $\Phi$ found in Ref. \cite{MBMDeA} is
\begin{equation}
\Phi = \frac{1}{c^2}\, U + \frac{1}{2}\, \Psi.
\end{equation}
The solutions found for $\Phi$ and $\Psi$ define the space-time metric (\ref{metric}).

%%%%%%%%%%%%%%%%%%%%%%%%%%%%%%%%%%%%%%%%%%%%%%%%%%%
\subsection{Fifth force on Earth and Moon}
%%%%%%%%%%%%%%%%%%%%%%%%%%%%%%%%%%%%%%%%%%%%%%%%%%%

In this section we focus on the equations of motion of Earth and Moon in the gravitational field of the Sun found in Ref. \cite{MBMDeA}. The equations describing the dynamics of the system
are obtained by taking the covariant divergence of the energy-momentum tensor and applying Bianchi identities to the gravitational field equations (see Ref. \cite{BBHL}),
as given by Eq. (\ref{covar-div-1}) that we repeat for convenience:
\begin{equation}
\nabla_\alpha T^{\alpha\beta} = \frac{f^2_R }{ 1 + f^2} ( g^{\alpha\beta} \LL_m - T^{\alpha\beta} ) \nabla_\alpha R.
\end{equation}
Thus, the computation executed in Ref. \cite{MBMDeA} gives the following equations of NMC dynamics of continuous bodies in hydrostatic state of stress and in the nonrelativistic limit:
\begin{equation}\label{NMC-contin-bodies}
\rho\frac{d\vv}{dt} = \rho\nabla U - \nabla p - \frac{1}{2} \rho c^2 \nabla\Psi - c^2 f^2_R\,\rho\nabla R,
\end{equation}
where the vector notation has been used. These equations are the Eulerian equations of Newtonian hydrodynamics with the presence of two additional terms:
\begin{itemize}
\item[{\rm (i)}] a fifth force density proportional to the gradient of the metric potential $\Psi$;
\item[{\rm (ii)}] an extra force density proportional to the product of $f^2_R$ by the gradient of curvature $R$.
\end{itemize}
The extra force density in (ii) has been extensively discussed in Ref. \cite{BBHL},
and for relativistic perfect fluids in Ref. \cite{BLP}.
While the fifth force is typical of $f(R)$ gravity theory, the extra force is specific of NMC gravity for the choice Eq. (\ref{f1-f2-specific}).
In Ref. \cite{MBMDeA} it has been found that the extra force is negligible in the Sun-Earth-Moon system, though it is expected to become significative at the galactic scale,
hence it will not be considered in the sequel.

Now we consider the motion of centers of mass of Earth and Moon. By using the continuity equation and Reynolds transport theorem of continuum mechanics, we have
\begin{equation}\label{Rey-mass-center}
M_\oplus\frac{d^2\x_E}{dt^2} = \int_{\VV_E(t)} \frac{d\vv}{dt}\rho_E(\x,t)d^3x,
\end{equation}
where $M_\oplus$ is the mass of Earth, $\VV_E(t)$ is the region of space occupied by Earth at the time instant $t$, and an analogous equation holds for the Moon.
Then, substituting the expression of $\rho d\vv/dt$ given by Eq. (\ref{NMC-contin-bodies}) into the integral in Eq. (\ref{Rey-mass-center}),
and neglecting the contribution from the extra force, we obtain
\begin{equation}\label{total-Earth-accelerat}
M_\oplus\frac{d^2\x_E}{dt^2} = \int_{\VV_E(t)}\rho_E\nabla U\, d^3x - \frac{c^2}{2} \int_{\VV_E(t)}\rho_E\nabla\Psi\, d^3x,
\end{equation}
where atmospheric pressure has been assumed uniform on the Earth surface.
The first integral is the contribution of Newtonian gravity to the acceleration of Earth, while the second integral is the contribution of fifth force. An analogous formula holds for the Moon.

In Ref. \cite{MBMDeA} it has been found that the contribution to the fifth force from the interior of the screening sphere is negligible due to the smallness of $\lambda(\rho_E)$, 
so that the contribution to the integral over $\VV_E(t)$ only comes from the thin shell defined by $r_E < \vert\x-\x_E\vert < R_\oplus$.
The integral over the thin shell is evaluated by using the expression (\ref{potential-Psi-sol}) of the potential $\Psi$ in terms of $\eta$,
which gives $\nabla\Psi=-(8\pi G/c^4)\nabla\eta$, and using the solution Eq. (\ref{eta-solut-Green}) for the function $\eta$.

In the following we denote $\x_S,\x_M$ the position vectors of the centers of Sun and Moon, we denote $M_\odot,M_M$ the masses of Sun and Moon,
and we denote $M_{\odot,{\rm eff}},M_{M,{\rm eff}}$ the respective effective masses. The contribution from the solar term $\eta_S$ is
\begin{eqnarray}
\frac{4\pi G}{c^2}\int_{\VV_E(t)} \rho_E\nabla\eta_S d^3x &=& \frac{G}{3}M_{\odot,{\rm eff}}M_{\oplus,{\rm eff}}\frac{\x_S-\x_E}{\vert\x_S-\x_E\vert^3} \nonumber\\
&+& \OO\left(\frac{R}{d}\right) \,,
\end{eqnarray}
and the contribution from the lunar term $\eta_M$ is analogous. Eventually, the contribution from the terrestrial term $\eta_E$ is
\begin{equation}\label{Earth-first-integral}
\int_{\VV_E(t)} \rho_E\nabla\eta_Ed^3x =  \OO\left(\frac{R}{d}\right).
\end{equation}
It turns out that all contributions to the fifth force of the order of $R\slash d$ cancel each other \cite{MBMDeA}. Since in the Sun-Earth-Moon system the ratios $R\slash d$ are small, then
corrections of the order of $(R\slash d)^2$ can be neglected because exceedingly small to give rise to observable effects.

Combining all these results, the acceleration of the Earth due to fifth force is given by
\begin{eqnarray}\label{Earth-fifth-force-result}
M_\oplus\left(\frac{d^2\x_E}{dt^2}\right)_{\rm f} &=& \frac{G}{3}M_{\oplus,{\rm eff}}\left[ M_{\odot,{\rm eff}}\frac{\x_S-\x_E}{\vert\x_S-\x_E\vert^3} \right. \nonumber\\
&+& \left. M_{M,{\rm eff}}\frac{\x_M-\x_E}{\vert\x_M-\x_E\vert^3} \right].
\end{eqnarray}
The acceleration of the Moon due to fifth force is analogous:
\begin{eqnarray}\label{Moon-fifth-force-result}
M_M\left(\frac{d^2\x_M}{dt^2}\right)_{\rm f} &=& \frac{G}{3}M_{M,{\rm eff}}\left[ M_{\odot,{\rm eff}}\frac{\x_S-\x_M}{\vert\x_S-\x_M\vert^3} \right. 
\nonumber\\
&+& \left. M_{\oplus,{\rm eff}}\frac{\x_E-\x_M}{\vert\x_E-\x_M\vert^3} \right].
\end{eqnarray}
In the case of astronomical bodies with uniform mass density the above expressions coincide with the accelerations of Earth and Moon found in \cite{KW} for chameleon gravity and in
\cite{Capoz-Tsu} for $f(R)$ gravity theory. In the more realistic case of bodies with varying density such expressions give different results.
Moreover, the integral equations that determine the screening radii in NMC gravity are, in general, different from the corresponding ones in $f(R)$ gravity \cite{MBMGDeA,Burrage}.

%%%%%%%%%%%%%%%%%%%%%%%%%%%%%%%%%%%%%%%%%%%%%%%%%%%%%%%%%%
\subsection{Weak Equivalence Principle violation}
%%%%%%%%%%%%%%%%%%%%%%%%%%%%%%%%%%%%%%%%%%%%%%%%%%%%%%%%%%

Since the accelerations of the Earth and Moon due to fifth force depend on the effective masses of the bodies, which are the masses of the respective thin shells
and these depend on the internal structure of the bodies through density and size, then the Earth and Moon fall towards the Sun with different accelerations.
Hence, a violation of the universality of free fall (UFF) takes place.

Using the expressions of the acceleration of Earth and Moon due to fifth force, the leading terms of the relative Earth-Moon acceleration are given by
\begin{eqnarray}
\a_M - \a_\oplus &=& -GM^\ast\frac{\x_M-\x_E}{\vert\x_M-\x_E\vert^3} \nonumber\\&+& \Delta_{\rm ESM}GM_\odot\frac{\x_S-\x_E}{\vert\x_S-\x_E\vert^3} 
 \nonumber\\&+& \mbox{tidal terms},
\end{eqnarray}
where
\begin{eqnarray}
\Delta_{\rm ESM} &=& \frac{1}{3}\frac{M_{\odot,{\rm eff}}}{M_\odot}\left( \frac{M_{M,{\rm eff}}}{M_M} - \frac{M_{\oplus,{\rm eff}}}{M_\oplus} \right), \label{Delta-ESM}\\
M^\ast &=& \left(M_\oplus + M_M\right)\left(1 + \frac{1}{3}\frac{M_{\oplus,{\rm eff}}}{M_\oplus}\frac{M_{M,{\rm eff}}}{M_M}\right).
\end{eqnarray}
The meaning of the terms in the expression of $\a_M-\a_\oplus$ is the following \cite{Wi}:
\begin{itemize}
\item[{\rm (i)}] the first term is the relative acceleration due to the gravitational attraction between the Earth and Moon;
\item[{\rm (ii)}] the second term, which can be written in the form,
\begin{equation}\label{UFF-accell}
\Delta_{\rm ESM}\, \mathbf{g}_S,
\end{equation}
$\mathbf{g}_S$ being the Newtonian acceleration of Earth due to the Sun, is the UFF violation-related difference between the Earth and the Moon accelerations towards the Sun, hence,
in the framework of NMC gravity, this term gives rise to a violation of the WEP.
\end{itemize}
The size of the UFF violation is represented by the parameter $\Delta_{\rm ESM}$, where ESM stands for Sun, Earth and Moon, since, by definition of the effective mass,
such a parameter depends on the composition (density) and size of all the three astronomical bodies \cite{Visw-Fienga}. Particularly, the WEP violation depends on
size and composition of the Sun, in addition to the more usual dependence on size and composition of the Earth and Moon.

If the astronomical bodies are screened with screening radii close enough to radii of the bodies, then
the effective masses and, consequently, $\Delta_{\rm ESM}$ can be made small enough in such a way that an experimental bound on WEP can be satisfied.

%%%%%%%%%%%%%%%%%%%%%%%%%%%%%%%%%%%%%%%%%%%%%%%%%%%%
\section{Constraints from Cosmological Data}\label{sec:Constraints_CosmologicalData}
%%%%%%%%%%%%%%%%%%%%%%%%%%%%%%%%%%%%%%%%%%%%%%%%%%%%
We constrain the modified theory's parameter space using a Markov Chain Monte Carlo (MCMC) sampler in the \textsc{cobaya} package \cite{COBAYA_paper} and present the posterior contour plots using the \textsc{GetDist} package \cite{Lewis:2019xzd}. The chain convergence was assessed via the generalized Gelman-Rubin statistic built into the package, for which we used $(R_{\rm stat}-1)=0.03$. Priors for the density parameters $\Lambda$ and $\rho_0$ were chosen to be uniform over adequate ranges to ensure no bias was introduced,  allowing for $H_0$ to vary between $\sim50-80 \ $ km/s/Mpc and $\Omega_m$ to vary between $\sim0.1-0.5$. Note that we do not use $\Omega_m$ and $H_0$ as input variables, as we have shown in Eq. \eqref{eq:HubbleEquation} that in general for the tracking solution it is not possible to analytically write $\Omega_\Lambda$ as a function of $\Omega_m$ and thus cannot eliminate it in favor of $H_0$. We therefore use $\Lambda$ and $\rho_0$ as the input for the density parameter space, only transforming to $\Omega_m$ and $H_0$ at the posterior level after the completion of the MCMC chains.

The NMC coupling parameter $\mu$ is given a uniform prior that allows it to vary between GR ($\mu=0$) and values beyond the tracking and existence conditions \eqref{mu-bound-track} and \eqref{eq:MaxMu_ExistenceOfMin}, given the choice of $\eps=10^{-2}$. At each step in the chain, if $\mu$ takes a value that does not satisfy these conditions, then the likelihood of that point is set to 0, such that we ensure compliance with the tracking solution. It is worth noting that the model allows for larger values of $\mu$, but these cannot be analyzed in the same analytic manner as done in this work and therefore are outside of the scope of our present investigation. In what follows, we describe the data used in this analysis.

\subsection{DESY5}\label{subsec:DESYR5}
The Dark Energy Survey Year 5 (DESY5) sample\footnote{Data available at \url{https://github.com/des-science/DES-SN5YR}} consists of 1635 photometrically-classified type IA supernovae (SNIa) in the redshift range $0.1 <z< 1.3$ and complemented by 194 low-redshift SNIa in the range $0.025 < z < 0.1$ taken from historical sources, with some overlap with other samples, such as the Pantheon+ \cite{Scolnic:2021amr} and Union3 datasets. The DESY5 sample provides distance moduli $\mu(z)$ of SNIa, which may also be calculated theoretically from the cosmic expansion history as 
\begin{equation}\label{eq:DistanceModulusDefinition}
    \mu_d(z)=5\log_{10}\left(\frac{d_L(z)}{1 \ \text{Mpc}}\right)+25 \, ,
\end{equation}
where
\begin{equation}\label{eq:LuminosityDistance}
    d_L(z)=(1+z)c\int_0^z\frac{{\rm d}z'}{H(z')} \, .
\end{equation}
This dataset does not include calibration for the SNIa absolute magnitude $M_B$, in contrast to the Cepheid-calibrated Pantheon+ sample, thus being incapable of breaking the degeneracy between the Hubble constant $H_0$ and $M_B$. However, in the context of analyzing the dynamical nature of dark energy, one can marginalize over the parameter generating this degeneracy $\mathcal{M}=M_B+5\log_{10}(c/H_0)$ when calculating the $\chi^2$ value for each point in the parameter space \cite{DES:2024fdw}, as only the shape of the evolution $E(z)\equiv H(z)/H_0$ is relevant. This is done by defining 
\begin{equation}\label{eq:SNIa_Marginalize}
    \tilde\chi^2_{\rm SNIa}=\chi^2_{\rm SNIa}-\frac{B^2}{c}+\ln\left(\frac{C}{2\pi}\right)\, ,
\end{equation}
where 
\begin{equation}
B=\sum_i(\mathbf{C}^{-1}_{\text{stat+sys}}\Delta\vec{\mathbf{   D}})_i
\end{equation}
and
\begin{equation}
C=\sum_i\sum_j\left[\mathbf{C}^{-1}_{\text{stat+sys}}\right]_{ij}
\end{equation}
are defined in terms of the difference between theory and observation $\Delta{\vec{\mathbf{D}}}_i=\mu_{d,i}-\mu_{d,\rm th}(z_i)$ and the full systematic and statistical covariance matrix $\mathbf{C}_{\text{stat+sys}}$, identically to their original calculation in Ref. \cite{Goliath:2001af}.

\subsection{Pantheon+}
The Pantheon+ dataset\footnote{Data available at \url{https://github.com/PantheonPlusSH0ES/DataRelease}} consists of a sample of 1701 cosmologically viable SNIa light curves from 1550 distinct supernovae in the redshift range $0.001<z<2.26$ \cite{Scolnic:2021amr}. It is a combination of 3 distinct mid-$z$ samples ($0.1<z<1.0$), 11 separate low-$z$ samples ($z<0.1$) and 4 different high-$z$ samples ($z>1.0$), each with their unique photometric systems and selection functions. Similarly to the DESY5 data, Pantheon+ provides measurements of the distance moduli $\mu(z_n)$. The degeneracy between $H_0$ and $M_B$ may be relaxed by including data from SH0ES Cepheid host distance anchors \cite{Riess:2021jrx}, made available in the Pantheon+SH0ES data release. However, in the context of the dynamical nature of dark energy, regardless of it emerging from a modified matter or gravity sector, the impact of a precise measurement of the Hubble constant or the absolute magnitude of the SNIa is negligible, as we are uniquely concerned with the relative shape of the evolution of the expansion rate and not its absolute value. In the spirit of efficiency of our MCMC analysis, we therefore follow the same methodology as for the DESY5 data to reduce our parameter space by marginalizing over these two parameters using Eq. \eqref{eq:SNIa_Marginalize}. This is done using the corresponding Pantheon+ covariance matrix, provided by the collaboration along with the distance moduli values.

\subsection{DESI DR2 BAO}\label{subsec:DESIDR2}
The distance over which baryon acoustic oscillations (BAO) propagate in the primordial fluid in the early Universe is fixed by the sound horizon at baryon decoupling $(z_d\simeq1089)$ and leaves an observable imprint in the matter distribution, which we may experimentally measure in the late-time distribution of galaxies. In their second data release (DR2), the Dark Energy Spectroscopic Instrument (DESI) collaboration provides BAO measurements in seven redshift bins from more than 30 million galaxies and quasars, and Ly$\alpha$ forest spectra of more than 820,000 quasars. The scale set by BAO serves as a cosmological standard ruler, as this distance is defined by the sound horizon as the time of baryon-photon decoupling
\begin{equation}
    r_d=\int_{z_d}^\infty\frac{c_s(z)}{H(z)}dz \,,
\end{equation}
where the speed of sound in the photon-baryon fluid can be determined as
\begin{equation}
    c_s(z)=\frac{c}{\sqrt{3\left(1+\frac{3\rho_b(z)}{4\rho_\gamma(z)}\right)}} \, ,
\end{equation}
with $\rho_b$ and $\rho_\gamma$ denoting the baryon and photon density, respectively. Since the integration domain ranges from the baryon decoupling redshift to infinity, the sound horizon is independent of modified gravity effects in the late-Universe. 

Measurements of the BAO scale in the transverse direction constrain the transverse comoving distance, given in a spatially flat Universe $(\Omega_k=0)$ by
\begin{equation}
    D_M(z)=\frac{c}{H_0}\int^z_0\frac{dz'}{E(z')}\,.
\end{equation}
Similarly, a measurement in the line-of-sight direction directly constrains the expansion rate via the distance
\begin{equation}
    D_H(z)=\frac{c}{H(z)}\,.
\end{equation}
For redshift bins with low signal-to-noise ratio, a measurement of the dilation scale 
\begin{equation}
    D_V(z)=\left[zD_M(z)D_H(z)\right]^{1/3}
\end{equation}
is given. As the inferred distances are determined in relation to the sound horizon, the constrained quantities are the ratios $D_M/r_d$, $D_H/r_d$ and $D_V/r_d$. As all of the distances are inversely proportional to the Hubble constant $D\propto H_0^{-1}$, the BAO measurements have a degeneracy between this quantity and the sound horizon at decoupling, effectively only constraining the combination $H_0r_d$, unless one breaks this degeneracy with other constraints on the sound horizon. 

Similarly to the work on the inverse distance ladder by the DES collaboration \cite{DES:2024ywx}, where the SNIa data were combined with DESI BAO data to break the degeneracy between $H_0$ and $M_B$, we use $r_d=147.46 \ \rm Mpc$, as determined in Ref. \cite{Lemos:2023xhs} from CMB constraints on the early Universe independently of late-time cosmology. In fact, although in what follows we determine the value of the Hubble constant from the constraints on the NMC model, this value has a degeneracy with $r_d$, as expected due to DESI providing measurements of $D_i/r_d\propto(H_0r_d)^{-1}$. This should be kept in mind when considering our posterior for $H_0$. Regardless, it has no direct effect on the DDE behaviour of the NMC theory, as this only depends on the ratio $\chi=4\Omega_\Lambda/\Omega_m$ and on the coupling parameter $\mu$, such that fixing the sound horizon does not impact the final conclusions drawn from our investigation.

The $\chi^2$ value for the BAO measurements is calculated using the covariance matrix made available along other values in the DESI second data release\footnote{Data available at \url{https://github.com/CobayaSampler/bao_data/tree/master/desi_bao_dr2}}.

\subsection{Constraints on tracking solution}\label{subsec:Constraints_Cosmology}
\begin{figure}[t!]
    \centering
    \includegraphics[width=\linewidth]{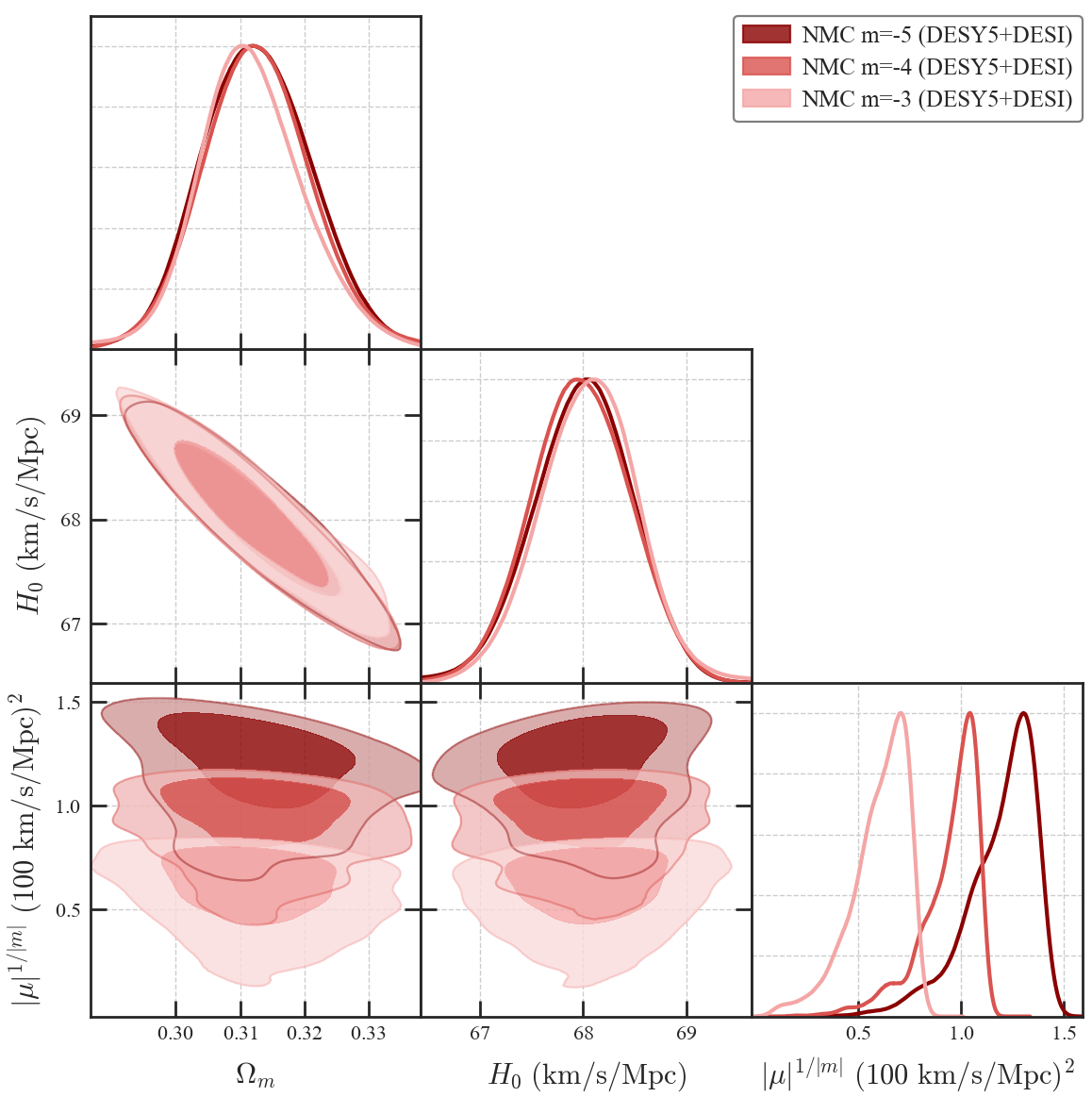}
    \includegraphics[width=\linewidth]{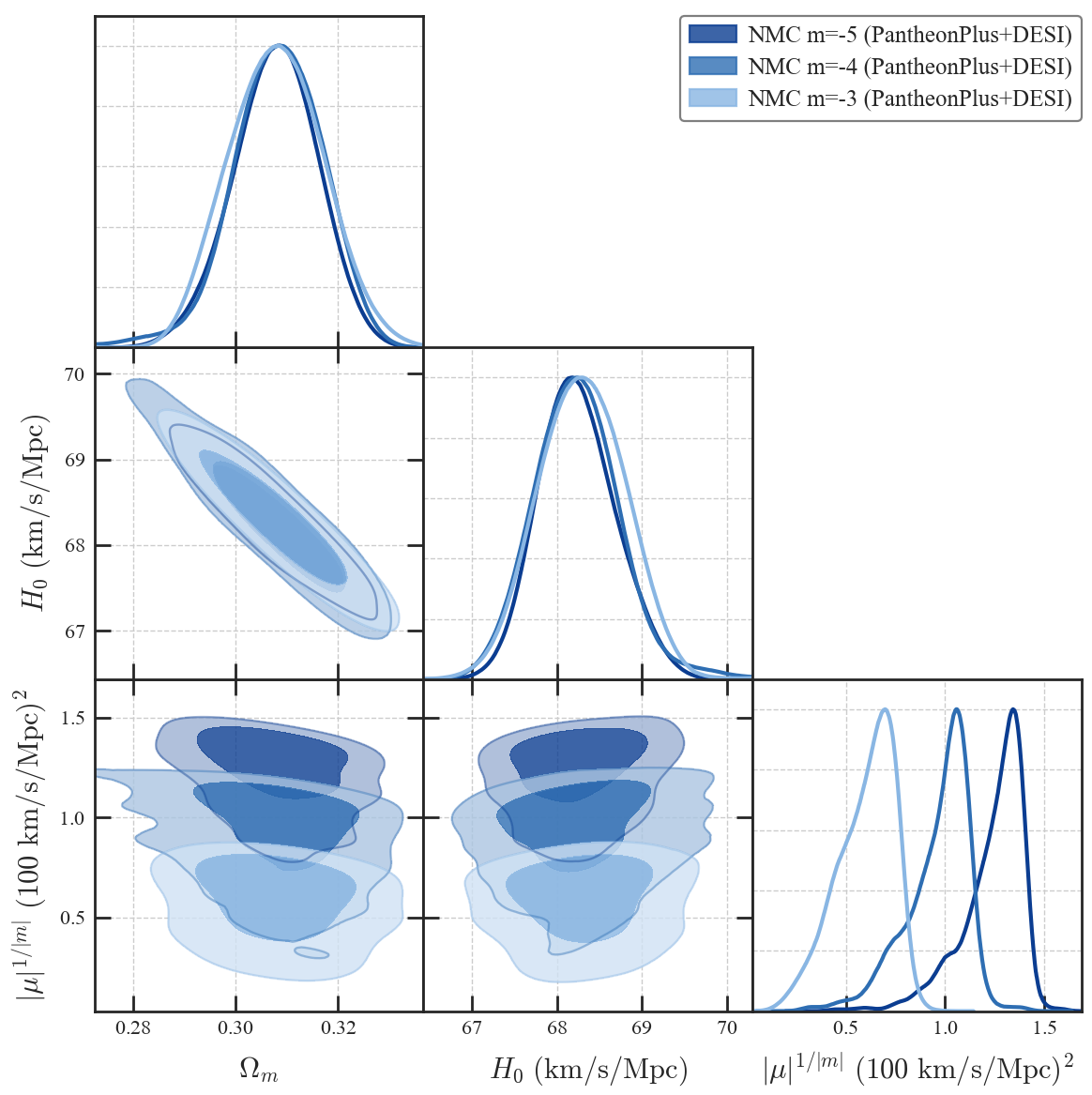}
    \caption{Constraints on cosmological parameters for the NMC models with different values of exponent $m$ in $f^2(R)$. All models agree on identical values of $\Omega_m$ and $H_0$, with the latter being related to the choice of fixing the sound horizon, with which it is degenerate. This has no effect on the NMC coupling, which grows with increasing $|m|$, as expected from the suppression of NMC effects associated with larger exponents.}
    \label{fig:BackgroundPosteriors}
\end{figure}

The constraints on cosmological parameters from DES/Pantheon+ SNIa and DESI BAO data are shown in Fig. \ref{fig:BackgroundPosteriors}. Notably, the posteriors for the matter density parameter $\Omega_m$ and the Hubble constant $H_0$ are mostly independent of the exponent $m$. Overall, all NMC models analyzed here point to similar results of $\Omega_m\approx0.313\pm0.009$ and $H_0\approx(67.98\pm0.53) \ {\rm km/s/Mpc}$, both within $1\sigma$ of the DESI DR2 results \cite{DESI:2025zgx}, which used Big Bang nucleosynthesis (BBN) priors. This is also in agreement with the separate MCMC analysis we have conducted for the $\Lambda$CDM model as a baseline comparison for the NMC tracking solution. The $H_0$ value aligning with that of the Planck 2018 results \cite{Planck:2018vyg} is expected, as we have fixed the sound horizon from CMB constraints on the early Universe \cite{Lemos:2023xhs}. As previously mentioned, this has no direct consequence on the DDE behavior, which focuses on the relative evolution of dark energy rather than the absolute value of the dust and dark energy densities. We also find that the results from DESY5 and Pantheon+ are effectively equivalent, such that no differing conclusions can be drawn by analyzing one dataset over another.

We present the constraints on the NMC parameter $\mu$ in terms of the characteristic curvature scale associated with each exponent $m$, defined as $R_c\equiv|\mu|^{1/|m|}$, here presented in the same units as the Hubble parameter. As shown in Fig. \ref{fig:BackgroundPosteriors}, the best-fit value for this curvature scale grows for larger values of $|m|$. This is expected, as for larger exponents the effects of the tracking solution are more suppressed, thus requiring larger values of $|\mu|$ to ensure we obtain similar cosmological effects. Interestingly, all of the best-fit values of $\mu$ differ by more than $1\sigma$ from the GR value $\mu=0$. This indicates that the dynamical dark energy behavior introduced by the tracking solution is preferred by cosmic expansion history data, confirming the trend that there seems to be more to dark energy than a cosmological constant. 

Given our constraints for $\mu$, $\Omega_m$ and $H_0$, we may calculate the value of $\xi_{\rm max}$ using Eqs. \eqref{eq:xi_max} and \eqref{relat-mu-xi0-approx}. Taking the example of $m=-4$, we find that the maximum $\xi$ value occurs at $z\sim0.35$, peaking at $\xi_{\rm max}\approx2\times10^{-3}$. This confirms that our approximation $\xi\ll1$ is self-consistent when imposing $\varepsilon=10^{-2}$, as we did in all numerical calculations for the data analysis.

However, focusing on the DESY5+DESI compilation and looking at the $\chi^2$ for each exponent $m$, we find that at best this value is $\chi^2=1695$ for $4\lesssim|m|\lesssim5$. In comparison, the $\Lambda$CDM model fits the data with one less parameter ($\mu$) and obtains $\chi^2=1700$. Although this is technically larger than that of the $m=-4$ and $m=-5$ NMC models, it is important to note that this does not account for the additional parameter used in the latter. Therefore, we calculate the values of the Akaike Information Criterion (AIC) \cite{AIC} and Bayesian Information Criterion (BIC) \cite{BIC}, defined respectively by
\begin{align}
    AIC&=2n-2\ln\mathcal{L}^{\rm max} \, ,\\
    BIC&=n\ln N-2\ln\mathcal{L}^{\rm max} \, ,
\end{align}
where $n$ is the number of fitted parameters in the model, $\mathcal{L}^{\rm max}$ is the maximum posterior likelihood from the MCMC analysis and $N$ is the number of data points in the analyzed sample. As suggested in Ref. \cite{AIC_Evidence}, relative values of $\Delta AIC>2,\ 5$ and 10 indicate weak, moderate and strong evidence against the model with the higher $AIC$ value, and equivalently for $BIC$. Due to the large amount of data points ($\ln N\sim8>2$), the $BIC$ value penalizes the usage of additional parameters, as in the case of the NMC models in this work, more severely than its counterpart ($AIC$). For the best case of the $m=-4$ and $m=-5$ models, the values in comparison with $\Lambda$CDM are $\Delta AIC=AIC_{\Lambda\rm CDM}-AIC_{\rm NMC}\approx+3$ and $\Delta BIC\approx-3$, indicating weak preference for NMC and $\Lambda$CDM respectively. This comes down to the weight of the penalty assigned to the number of fitted parameters. This shows that even with an additional parameter, the best NMC models are not conclusively superior to $\Lambda$CDM. 

Nevertheless, there is an interesting conclusion to be drawn from our results. All NMC models were allowed to vary between strong couplings (within the validity of the tracking solution) and no coupling whatsoever ($\mu=0$). Even with this freedom, the data still led us to the presence of a non-minimal coupling which infects the cosmological constant with DDE-like behavior, even if with less statistical significance than what was reported by the DESI collaboration by using the $w_0w_a$ parametrisation \cite{DESI:2024mwx,DESI:2025zgx}. 

\begin{figure}[t!]
    \centering
    \includegraphics[width=\linewidth]{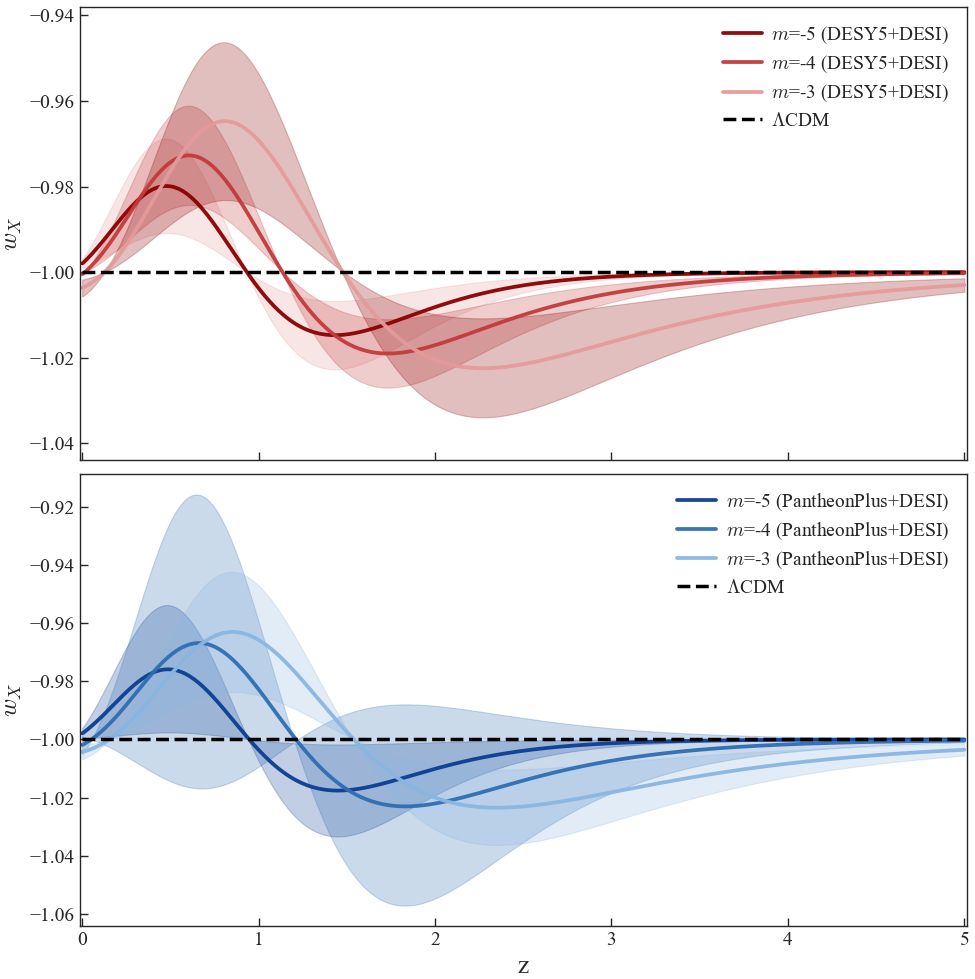}
    \caption{Constraints on the effective dark energy equation of state parameter $w_X$ for the best fit of the NMC models with $|m|=\{3,4,5\}$ to the DESY5+DESI and Pantheon+DESI data compilations. Each shaded region identifies the corresponding $1\sigma$ region for each model. All models exhibit at least one phantom crossing at $z\sim1-2$.}
    \label{fig:wX_BestFit}
\end{figure}

The qualitative behavior of the effective equation of state parameter of dark energy for the best-fit values of the NMC models with best statistical performance is similar. As shown in Fig. \ref{fig:wX_BestFit}, all models exhibit at least one phantom crossing, {\em i.e.} $w_X$ crossing over to $w_X<-1$, with the crossing redshift $z_c$ becoming smaller for increasing $|m|$. This is expected, as increasing $|m|$ leads to the NMC effects being suppressed until later redshifts and having an overall weaker deviation from $\Lambda$CDM within the tracking condition. The expression Eq. (\ref{equat-state-first-order}) of $w_X$ shows that the crossing redshift $z_c$ is fixed for each $m$ by the ratio between $\Lambda$ and $\rho_0$, which is quite strictly fixed by the cosmic expansion history, thus explaining why the $1\sigma$ regions funnel when approaching this crossing. These differ from the DESI result of $z_c\sim0.5$ for the $w_0w_a$ parametrisation, while still providing a possible physical explanation for the emergence of a phantom behavior of dark energy. However, the $m=-3$ and $m=-4$ NMC models undergo a second phantom crossing closer to the present $z_c\sim0$, with the same expected for $m=-5$ for near-future redshifts by analogy. This very low redshift region is not probed by the presently limited BAO data points provided by the DESI collaboration. Future surveys may provide sufficiently varied and accurate data to decidedly constrain the nature of dark energy \cite{Miao:2023umi,Verdier:2025gak,Qin:2025nkk,Pan:2023zgb,Euclid:2025dlg,Ding:2023ibo}. 

\subsection{Comparison with LLR constraints}\label{subsec:ComparisonWithLLR}

We now aim to compare the constraints from cosmological expansion history determined in this section to those found from LLR data. The latter were determined in Ref. \cite{MBMDeA}, in which the slightly different notation for the NMC parameters $q\leftrightarrow\mu$ and $\alpha\leftrightarrow m$ was used. For consistency, we will refer to these quantities by using the notation used thus far in the present investigation. Moreover, in order to determine a dimensionless value for the NMC coupling scale constrained from LLR data, the parameter $\mu$ was rescaled by the curvature $R_g=8\pi G\rho_g/c^2$, calculated in terms of the galactic mass density in the solar neighborhood of the Milky Way $\rho_g\approx6.9\times10^{-24} \ {\rm g/cm^3}$. This defines the dimensionless quantity $\tilde{\mu}=\mu R_g^m$. 

However, in the spirit of this work, it is useful to define a new quantity $\hat\mu=\mu R_0^m$ in terms of the cosmological dust matter density $\rho_0$. As seen in Fig.  \ref{fig:BackgroundPosteriors}, all NMC models agree on the values of $\Omega_m$ and $H_0$, allowing us to calculate a global best-fit value of $\rho_0$, which can be used to rescale the LLR results from Ref. \cite{MBMDeA} as 
\begin{equation}
    \hat\mu=\tilde\mu\left(\frac{R_0}{R_g}\right)^m \,. 
\end{equation}
In Ref. \cite{MBMDeA} constraints on the parameters $m,\tilde{\mu}$ were found by resorting to a constraint in terms of difference between the Earth and the Moon accelerations towards the Sun given in Ref. \cite{Visw-Fienga}.
In order to test UFF violations, a supplementary acceleration of the form (\ref{UFF-accell}) is introduced in the geocentric equation of
motion of the Moon. The parameter $\Delta_{\rm ESM}$ is estimated in the LLR adjustment together with a set of parameters of the lunar ephemerides listed in \cite{Visw-Fienga}.
The result on the UFF violation parameter in \cite{Visw-Fienga}, based on 48 years of LLR data, is given by
\begin{equation}\label{Delta-ESM-estimate}
\Delta_{\rm ESM} = \left(3.8\pm 7.1\right)\times 10 ^{-14},
\end{equation}
with $3\sigma$ uncertainty.
Since in the expression (\ref{Delta-ESM}) for the WEP violation parameter $\Delta_{\rm ESM}$, using Eqs. (\ref{Earth-effective-mass}) and (\ref{Earth-screening-radius}),
the effective masses of Sun, Earth and Moon are functions of the screening radii of the bodies which depend on $m$ and $\tilde{\mu}$,
then in Ref. \cite{MBMDeA} the estimate (\ref{Delta-ESM-estimate}) has been translated into a constraint on NMC gravity parameters.

\begin{figure}[t!]
    \centering
    \includegraphics[width=\linewidth]{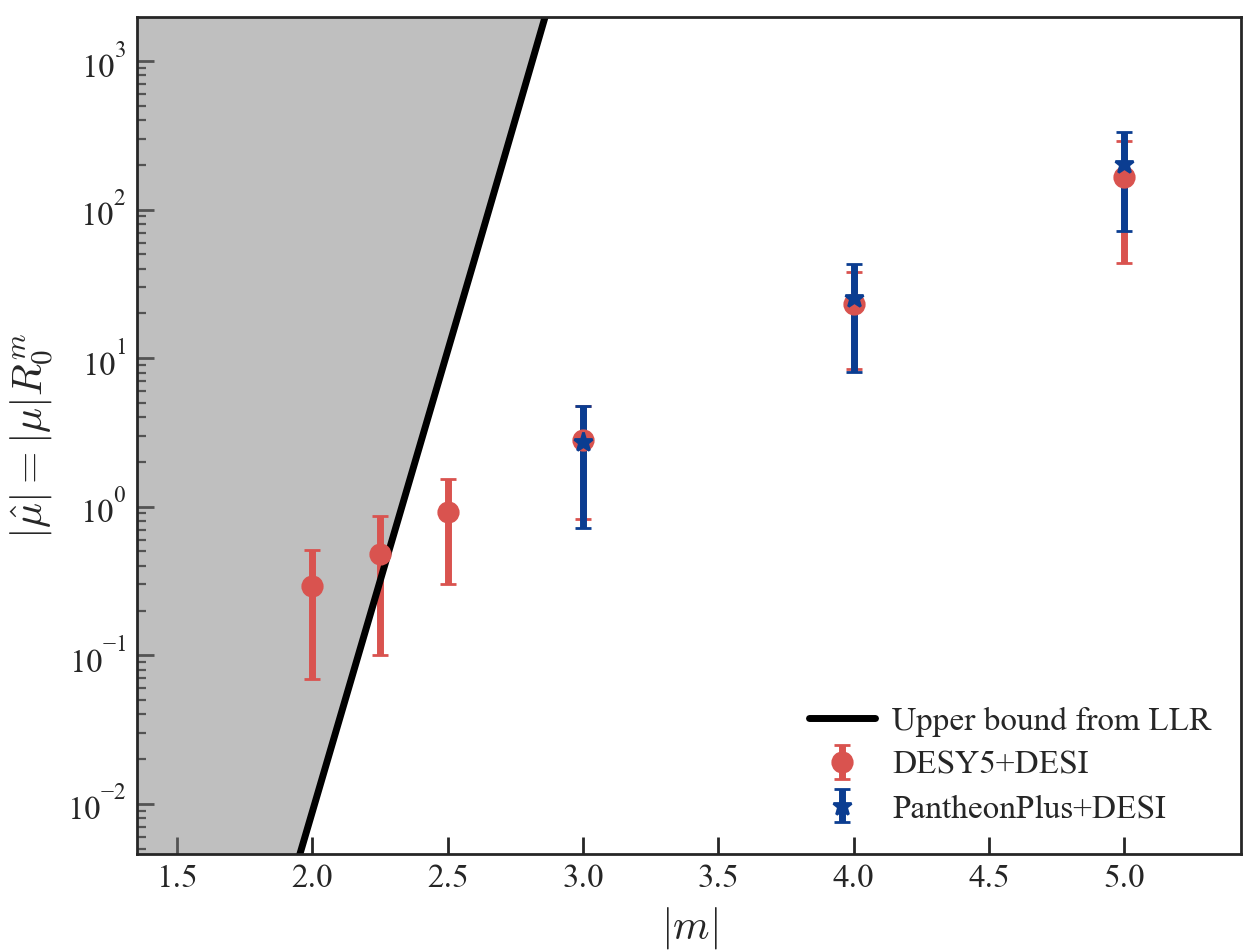}
    \caption{Constraints from cosmological data on the NMC parameters $\hat\mu$ and $m$ are shown in red (DESY5+DESI) and blue (Pantheon+DESI). The upper bound from LLR data is shown in black. NMC models with $|m|\lesssim2.25$ are ruled out by LLR constraints.}
    \label{fig:mu_m_Constraints}
\end{figure}

The constraints from Section \ref{subsec:Constraints_Cosmology} are shown in Fig. \ref{fig:mu_m_Constraints} together with the upper bound from LLR data computed in Ref. \cite{MBMDeA}. As expected, we find that the DESY5 and Pantheon+ constraints are practically indistinguishable and well within $1\sigma$ of each other. The best fit for the rescaled coupling parameter $|\hat\mu|$ grows with $|m|$ in an approximately exponential manner, as seen by the linear trend in the log-scale plot, although it does so at a slower rate than the upper bound on this parameter from LLR constraints. Our results show that models with $|m|\lesssim2.25$ are ruled out by the bounds from LLR data, while models with larger $|m|$ are able to accommodate cosmological data and still be consistent with the LLR upper limit. Considering that $|m|=4$ and 5 yielded the best results for the Pantheon+/DES and DESI data compilations, this points to the phenomenological viability of these models. It is important to note that the tracking condition imposed explicitly in our analysis limits the magnitude of the $|\hat\mu|$ parameter via Eqs. \eqref{eq:MaxMu_Tracking} and \eqref{eq:MaxMu_ExistenceOfMin}. Although it would force a reinvention of the analytical approach used in this investigation, removing this restriction in a future analysis would allow for stronger nonminimal couplings, which could in turn put the model in stronger tension with the LLR constraints.

The tightest constraint to date on WEP has been provided by the MICROSCOPE mission with a precision of $10^{-15}$ \cite{Touboul}. The WEP violation in the Earth-Moon system is due to the fact that the bodies are not completely screened by the chameleon mechanism. In Ref. \cite{Pernot} the data from the MICROSCOPE experiment were used to test chameleon theory of gravity. However, since this experiment was not designed for the purpose of the analysis of modified gravity theories of this type, the bounds on parameters of chameleon gravity found in Ref. \cite{Pernot} are not yet competitive with state-of-the-art constraints.

We conclude this section by comparing the above results with the ones achieved by a $f(R)$ gravity model which adds a cosmological constant to a linear function of $R$ plus a power law function.
Using our notation, we have $f^2(R)=0$ and $f(R)=f^1(R)$ with
\begin{equation}
f^1(R) = \frac{c^4}{8\pi G}(R-2\Lambda) + \mu R^m.
\end{equation}
This $f(R)$ model has been analyzed in Ref. \cite{FTBM}, where the authors consider values $0<m<1$ and $\mu<0$, which ensure stability both at the cosmological scale and at the Solar System scale. The results of Ref. \cite{FTBM} show that the Solar System bound on the PPN parameter $\gamma$ (although weaker than the LLR bound) constrains the effective equation of state parameter for dark energy to be
\begin{equation}
\vert w_X+1 \vert \lesssim 0.3\times 10^{-6},
\end{equation}
which is unobservable compared with DESI data. It turns out that Solar System constraints on cosmological amplitudes are strong in the class of $f(R)$ gravity models with $m>0$ \cite{HS}. Moreover, models with $m<0$ are ruled out because of their inherent instability \cite{FTBM}. Conversely, in the NMC gravity model, negative exponents $m$ yield a stable solution and Solar System constraints still allow an observable dynamic dark energy.

%%%%%%%%%%%%%%%%%%%%%%%%%%%%%%%%%%%%%%%%%%%%%%%%%%%%%%%%%%
\section{Conclusions}\label{sec:Conclusions}
%%%%%%%%%%%%%%%%%%%%%%%%%%%%%%%%%%%%%%%%%%%%%%%%%%%%%%%%%%
In this work, we have analyzed a modified theory of gravity with a curvature-dependent non-minimal coupling between matter and curvature of the form $f^2(R)$. After presenting the field and the conservation equations, we discussed our choice of a NMC function that evolves as an inverse power-law of the Ricci scalar $f^2(R)=\mu R^m$, with $\mu<0$ fixed by the existence and stability of the chameleon solution in the Solar System, while $m<0$ ensures the emergence of NMC effects in the late Universe. We then introduced a scalar field $\eta$ such that the trace of the field equations can be recast as describing $\eta$ moving in an effective potential. This NMC is applied to a homogeneous and isotropic Universe with a positive cosmological constant, such that the modified theory's effect is to drive a dynamical behavior of dark energy motivated by the latest DESI results.  

We then investigated the properties of this effective potential, finding a condition for the existence of a minimum from the beginning of the matter-dominated epoch in terms of the theory's parameters and the dark energy content of the Universe. Once this was established, we looked for a condition such that, for a small enough period of oscillations (relatively to the expansion rate of the Universe) around the minimum of the potential, one can consider the field $\eta$ to evolve according to the position of this minimum. The validity of this tracking solution was determined, thus adding a second constraint that must be obeyed, this time in terms of the model's parameters and the non-relativistic matter content of the Universe. We showed that the tracking condition is always stronger than that of the existence of a minimum of the effective potential, such that we were only required to check one of these in the posterior analyses in this work. Under these assumptions, we analytically calculated the impact of the NMC on the evolution of the effective dark energy content of the Universe, considered here to be a combination of $\Lambda$ and curvature-dependent terms.

Considering all these conditions, we reviewed past results from Ref. \cite{MBMDeA} on the effects of the NMC model on the dynamics of the Earth-Moon system, which are reflected as the
presence of a fifth force which, without a suitable screening mechanism, implies that the Earth and Moon fall differently towards the Sun, giving rise to a violation of the weak equivalence principle.

 Then we presented the cosmological datasets with which we have constrained the model. Our results indicate that combining DESI BAO data separately with Pantheon+ and DESY5 supernovae distance moduli samples leads to similar constraints on theory and background cosmology, with DESY5 providing smaller uncertainties due to the increased number of data points relatively to Pantheon+. Although models with $|m|=4$ and 5 obtained smaller $\chi^2$ values than $\Lambda$CDM, this is achieved by introducing the additional fit parameter $\mu$, which depending on the chosen criteria used to assess the quality of the fit leads to indecisive conclusions on the preference of the NMC model over $\Lambda$CDM. Nevertheless, the NMC models exhibit some qualitative properties of the dark energy equation of state parameter $w_X$ favored by DESI data, mainly a phase
 of $w_X>-1$ at low redshifts followed by a phantom crossing to $w_X<-1$ at a higher redshift.

We concluded the discussion by considering the constraints imposed on the theory's parameters from a test of the equivalence principle based on Lunar Laser Ranging data,
then intersecting the cosmological expansion history and LLR constraints to determine the physical viability of the class of models under consideration. The result of this highly non-trivial set of constraints stipulate that models with $|m|\lesssim2.25$ are ruled out by the LLR upper bound on $|\mu|$ when considering the best-fit values from the DESI and DESY5 samples. The fact that $|m|=4$ and 5 models stand out under the scrutiny of the LLR bounds, further stresses their phenomenological adequacy as a suitable effective theory to describe gravity.

Improvements concerning LLR data are expected from lunar laser retroreflectors of next generation such as MoonLIGHT (Moon Laser Instrumentation for General Relativity High
accuracy Tests) developed by National Institute of Nuclear Physics-National Laboratories of Frascati (INFN-LNF), supported by European Space Agency (ESA),
Italian Space Agency (ASI), and by the China National Space Administration (CNSA) with AIR-CAS (Aerospace Information Research Institute - Chinese Academy of Sciences), and waiting for launch to the Moon through the Commercial Lunar Payload Services (CLPS) NASA program in 2026 and through the CNSA mission Chang’e-7 to the South Pole also in 2026
\cite{DellAgnello1, DellAgnello2}.

This work has focused on analyzing the NMC model with an inverse power-law of the curvature scalar $R$ under the assumption of tracking solution for the scalar field $\eta$. This was motivated by the ability to make analytical predictions on the effects of the modified theory, which allowed for a more numerically efficient and thus detailed statistical treatment of the theory's predictions on the cosmological level. A natural extension would be to relax the assumption of the tracking solution and consider
a larger set of values of parameters $m,\mu$ such that the scalar field is no longer bound to the minimum, so to avoid an
addition of a cosmological constant to drive the Universe's accelerated expansion. That will require the introduction of suitable mathematical tools to study
the more general behavior of the NMC model which will be the subject of future research. Moreover, 
the resulting ability of the model to reproduce the dynamical evolution of dark energy has to be compared with new LLR constraints that will come out from the deployment
on the Moon of lunar laser retroreflectors of next generation, such as the ones quoted above.
These developments match with the exciting prospect of future surveys with increased precision of measurements both at the SNIa and BAO probes of cosmic expansion history, which may hopefully allow for a deeper understanding of dark energy and its dynamical nature or otherwise, as well as more stringent tests of the NMC $f(R)$ theory under consideration.

Such tests could come at the cosmological background level, since the theory modifies the Friedmann equation \cite{Bertolami:2013uwl}, which, in the particular context of this work, is manifested through a dynamical behavior of dark energy. This is captured by the evolution of the equation of state parameter $w_X$, which can show significant deviations from $\Lambda$CDM, particularly when moving beyond the tracking condition, as discussed above. This effect, although in this work limited to a small preponderance within the conditions imposed by the tracking solution, imprints itself on all distance-based observational aspects of cosmology, such as supernovae distance moduli or angular distance measurements from BAO \cite{BarrosoVarela:2024htf,BarrosoVarela:2024ozs}. 

Alternatively, one could also probe the theory at the level of perturbations, since it modifies GR by introducing an effective form of the gravitational constant in the master equation for energy density fluctuations, analogous to what is found in minimally coupled $f(R)$ gravity, as well as adding an exclusively NMC effect in the form of an additional friction term that follows from the modified conservation equations. Both scenarios are thus expected to present unique phenomenological imprints through their higher-order nature and explicit dependence on the Lagrangian density of perfect fluids. These will be most significant for stronger matter-gravity couplings, as well as for short scales, due to the higher-order terms in the wavenumber $k$ in the equations governing the evolution of density perturbations. Their observational consequences can be tested through the same probes used for minimally coupled $f(R)$ theories \cite{Mirzatuny:2019dux,BarrosoVarela:2025zzv}. These include, for example, modifications to the connection between the redshift-space galaxy power spectrum and the power spectrum in real space, as well as signals in the late-time integrated Sachs-Wolfe (ISW) effect, both of which follows from the modified growth of density perturbations in the theory.

\section*{Acknowledgments}

The work of R.M. is partially supported, and the work of G.B. and S.DA. is fully supported, by 
INFN (Istituto Nazionale di Fisica Nucleare, Italy), as part of the MoonLIGHT-2 experiment in the framework 
of the research activities of the Commissione Scientifica Nazionale n. 2 (CSN2).
S.DA. and G.B. would like to thank ASI (Agenzia Spaziale Italiana) for the support granted to them with ASI-INFN Agreement n. 2019-15-HH.0 and ESA (European Space Agency) for the support granted to them with ESA-INFN Contract n. 4000133721/21/NL/CR.

The work of M.B.V. is supported by FCT (Fundação para a Ciência e Tecnologia, Portugal) through the grant 2024.00457.BD. The work of O.B. is partially supported by FCT (Fundação para a Ciência e Tecnologia, Portugal) through the project 2024.00252.CERN.

%%%%%%%%%%%%%%%%%%%%%%%%%%%%%%%%%%%%%%%%%%%%%%%%%%%%%%%%%%
\appendix
\section{Effective dark energy}\label{sec:Appendix_effectiveDarkEnergy}
%%%%%%%%%%%%%%%%%%%%%%%%%%%%%%%%%%%%%%%%%%%%%%%%%%%%%%%%%%

The complete dark energy effective equation of state is the following:
\begin{equation}
w_X = - 1+\frac{N_{w_X}}{D_{w_X}} ,
\end{equation}
where $N_{w_X}$ and $D_{w_X}$ are given by
\begin{eqnarray}
N_{w_X} &=& \sum_{k=0}^3 {(1+\chi a^3)^k} \sum_{j=1}^6 V_{k,j}(m) \xi^j \nonumber\\
D_{w_X} &=& \sum_{k=0}^3 {(1+\chi a^3)^k} \sum_{j=0}^6 W_{k,j}(m) \xi^j,
\end{eqnarray}
and the coefficients $V_{k,j}$ and $W_{k,j}$ are polynomial functions of the parameter $m$. The leading-order functions are reported in Appendix \ref{sec:Appendix_Coefficients}.

The effective dark energy density is
\begin{equation}
\rho_X c^2 = \frac{c^4\Lambda}{8\pi G}\,\frac{N}{D},
\end{equation}
where $N$ and $D$ are given by
\begin{eqnarray}
N &=& \sum_{k=0}^2 {(1+\chi a^3)^k} \sum_{j=0}^3 Y^{(N)}_{k,j}(m) \xi^j, \nonumber\\
D &=& \sum_{k=0}^2 {(1+\chi a^3)^k} \sum_{j=0}^3 Y^{(D)}_{k,j}(m) \xi^j,
\end{eqnarray}
and the coefficients $Y^{(N)}_{k,j},Y^{(D)}_{k,j}$ are polynomial functions of parameter $m$. These functions are reported in Appendix \ref{sec:Appendix_Coefficients}.

%%%%%%%%%%%%%%%%%%%%%%%%%%%%%%%%%%%%%%%%%%%%%%%%%%%%%%%%%%

\section{Effective dark energy coefficients}\label{sec:Appendix_Coefficients}
%%%%%%%%%%%%%%%%%%%%%%%%%%%%%%%%%%%%%%%%%%%%%%%%%%%%%%%%%%

\subsection{Coefficients of effective dark energy equation of state}
%%%%%%%%%%%%%%%%%%%%%%%%%%%%%%%%%%%%%%%%%%%
\subsubsection{First order coefficients}
%%%%%%%%%%%%%%%%%%%%%%%%%%%%%%%%%%%%%%%%%%%
First we give the coefficients that are involved in the leading-order approximation of the fraction $N_{w_X}\slash D_{w_X}$ with respect to $\xi$:
\begin{eqnarray}
V_{0,1}(m) &=& 9m(2-3m+m^2), \nonumber\\
V_{1,1}(m) &=& 3m(-10+9m+m^2), \nonumber\\
V_{2,1}(m) &=& 2m(2+5m), \nonumber\\
V_{3,1}(m) &=& 2(2m-1), \nonumber\\
W_{0,0}(m) &=& 0, \nonumber\\
W_{1,0}(m) &=& 0, \nonumber\\
W_{2,0}(m) &=& m, \nonumber\\
W_{3,0}(m) &=& -m.
\end{eqnarray}
%
%%%%%%%%%%%%%%%%%%%%%%%%%%%%%%%%%%%%%%%%%%%
\subsubsection{Second order coefficients}
%%%%%%%%%%%%%%%%%%%%%%%%%%%%%%%%%%%%%%%%%%%

The coefficients that are involved in the second order Taylor expansion of the fraction $N_{w_X}\slash D_{w_X}$ with respect to $\xi$ are the following:
\begin{eqnarray}
V_{0,2}(m) &=& 9(6m^3-17m^2+14m-3), \nonumber\\
V_{1,2}(m) &=& 3(14m^3+29m^2-62m+19), \nonumber\\
V_{2,2}(m) &=&2(4m^3+9m^2+19m-10) , \nonumber\\
V_{3,2}(m) &=& 2m^2+5m-3, \nonumber\\
W_{0,1}(m) &=& 0, \nonumber\\
W_{1,1}(m) &=& 15m(m-1), \nonumber\\
W_{2,1}(m) &=& \frac{1}{2}(25m-1), \nonumber\\
W_{3,1}(m) &=& \frac{1}{2}(-6m^2+m-3).
\end{eqnarray}
%

%%%%%%%%%%%%%%%%%%%%%%%%%%%%%%%%%%%%%%%%%%%
\subsection{Coefficients of effective dark energy density}
%%%%%%%%%%%%%%%%%%%%%%%%%%%%%%%%%%%%%%%%%%%

The coefficients of the numerator are the following:
\begin{eqnarray}
Y^{(N)}_{0,0}(m) &=& 0, \nonumber\\
Y^{(N)}_{1,0}(m) &=&4m^2, \nonumber\\
Y^{(N)}_{2,0}(m) &=& -4m^2, \nonumber\\
Y^{(N)}_{0,1}(m) &=& 48m^2(m-1), \nonumber\\
Y^{(N)}_{1,1}(m) &=&2m(2m^2+17m-1), \nonumber\\
Y^{(N)}_{2,1}(m) &=&-2m(2m^2-3m+3), \nonumber\\
Y^{(N)}_{0,2}(m) &=& 48m(m-1)(2m-1), \nonumber\\
Y^{(N)}_{1,2}(m) &=& m(2m-1)(39-6m), \nonumber\\
Y^{(N)}_{2,2}(m) &=& m(2m-1)(2m-3), \nonumber\\
Y^{(N)}_{0,3}(m) &=& 12m(2m-1)^2(m-1), \nonumber\\
Y^{(N)}_{1,3}(m) &=& -4(2m-1)^2(m-3), \nonumber\\
Y^{(N)}_{2,3}(m) &=& 0.
\end{eqnarray}
The coefficients of the denominator are the following:
\begin{eqnarray}
Y^{(D)}_{0,0}(m) &=&0 , \nonumber\\
Y^{(D)}_{1,0}(m) &=& 4m^2, \nonumber\\
Y^{(D)}_{2,0}(m) &=& -4m^2, \nonumber\\
Y^{(D)}_{0,1}(m) &=& 12m^2(m-1), \nonumber\\
Y^{(D)}_{1,1}(m) &=&-2m(4m^2-11m+1) , \nonumber\\
Y^{(D)}_{2,1}(m) &=& -2m(2m^2+5m-1), \nonumber\\
Y^{(D)}_{0,2}(m) &=& 12m(m-1)(2m-1), \nonumber\\
Y^{(D)}_{1,2}(m) &=& -3m(2m-1)(4m-7), \nonumber\\
Y^{(D)}_{2,2}(m) &=& -9m(2m-1), \nonumber\\
Y^{(D)}_{0,3}(m) &=& 3(2m-1)^2(m-1), \nonumber\\
Y^{(D)}_{1,3}(m) &=& (2m-1)^2(6-4m), \nonumber\\
Y^{(D)}_{2,3}(m) &=& (2m-1)^2(m-3).
\end{eqnarray}

\bibliography{References} 
%%%%%%%%%%%%%%%%%%%%%%%%%%%%%%%%%%%%%%%%%%%
% %\subsection{Coefficients of the effective dark energy density}
% %%%%%%%%%%%%%%%%%%%%%%%%%%%%%%%%%%%%%%%%%%%

\end{document}